\documentclass[preprint2]{proto}
\usepackage{times}
\newcommand{\refs}{\par\noindent\hangindent=1pc\hangafter=1}
\newcommand{\eg}{{\em e.g.}}

\newcommand{\mum}{\,$\mu$m}

\newcommand{\el}{136108 (2003~EL$_{61}$)}
\newcommand{\fy}{136472 (2005~FY$_9$)}
\newcommand{\Eris}{136199 Eris}
\newcommand{\Sedna}{90377 Sedna}
\newcommand{\met}{CH$_4$}
\newcommand{\nit}{N$_2$}
\newcommand{\Spitzer}{{\em Spitzer}}
\newcommand{\tnmd}{$^{\hbox{\footnotesize\,d\,}}$}
\newcommand{\tnme}{$^{\hbox{\footnotesize\,e\,}}$}
\newcommand{\tnmf}{$^{\hbox{\footnotesize\,f\,}}$}

\begin{document}

\title{\textbf{\LARGE Physical Properties of Kuiper Belt and Centaur Objects:\\
Constraints from Spitzer Space Telescope}    
{\small \\To Appear in: {\em The Solar System beyond Neptune} (M.A. Barucci et al., Eds.) U. Arizona Press, 2007} 
}

\author {\textbf{\large John Stansberry}}
\affil{\small\em University of Arizona}
\author {\textbf{\large Will Grundy}}
\affil{\small\em Lowell Observatory}
\author {\textbf{\large Mike Brown}}
\affil{\small\em California Institute of Technology}
\author {\textbf{\large Dale Cruikshank}}
\affil{\small\em NASA Ames Research Center}
\author {\textbf{\large John Spencer}}
\affil{\small\em Southwest Research Institute}
\author {\textbf{\large David Trilling}}
\affil{\small\em University of Arizona}
\author {\textbf{\large Jean-Luc Margot}}
\affil{\small\em Cornell University}

\begin{abstract}
\baselineskip = 11pt
\leftskip = 0.65in 
\rightskip = 0.65in
\parindent=1pc
{\small Detecting heat from minor planets in the outer solar system
is challenging, yet it is the most efficient means for constraining
the albedos and sizes of Kuiper Belt Objects (KBOs) and their progeny,
the Centaur objects. These physical parameters are critical, \eg, for
interpreting spectroscopic data, deriving densities from the masses of
binary systems, and predicting occultation tracks.  Here we summarize
{\em Spitzer Space Telescope} observations of 47 KBOs and Centaurs at
wavelengths near 24 and 70\mum.  We interpret the measurements using 
a variation of the Standard Thermal Model (STM) to derive the physical
properties (albedo and diameter) of the targets.  We also summarize
the results of other efforts to measure the albedos and sizes of KBOs
and Centaurs.  The three or four largest KBOs appear to constitute a
distinct class in terms of their albedos.  From our Spitzer results,
we find that the geometric albedo of KBOs and Centaurs is correlated
with perihelion distance (darker objects having smaller perihelia), and
that the albedos of KBOs (but not Centaurs) are correlated with size
(larger KBOs having higher albedos). We also find hints that albedo may be
correlated with with visible color (for Centaurs).  Interestingly, if the
color correlation is real, redder Centaurs appear to have higher albedos.
Finally, we briefly discuss the prospects for future thermal observations
of these primitive outer solar system objects.
\\~\\~}%leave this in to get the correct vertical space after the abstract 

\end{abstract}

\section{\textbf{INTRODUCTION}}

The physical properties of Kuiper Belt Objects (KBOs) remain poorly known
nearly 15 years after the discovery of (15760) 1992~QB$_1$ ({\em Jewitt
and Luu}, 1993). While KBOs can be discovered, their orbits determined,
and their visible-light colors measured (to some extent) using modest
telescopes, learning about fundamental properties such as size, mass,
albedo, and density remains challenging. Determining these properties
for a representative sample of TNOs is important for several reasons.
Estimating the total mass of material in the transneptunian region,
and relating visible magnitude frequency distributions to size- and
mass-frequency is uncertain, at best. Quantitative interpretation of
visible and infrared spectra is impossible without knowledge of the
albedo in those wavelength ranges. Size estimates, when coupled with
masses determined for binary KBOs (see {\em Noll et al.} chapter),
constrain the density, and hence internal composition and structure,
of these objects.  All of these objectives have important implications
for physical and chemical conditions in the outer proto-planetary nebula,
for the accretion of solid objects in the outer Solar System, and for the
collisional evolution of KBOs themselves. Of course, there is a relative
wealth of information about Pluto and Charon, the two longest known KBOs,
and we do not address their properties further here.

The Centaur objects, with orbits that cross those of one or more of
the giant planets, are thought to be the dynamical progeny of KBOs
({\em e.g. Levison and Duncan}, 1997; {\em Dones et al.} chapter).
The Centaurs are particularly interesting both because of their direct
relation to KBOs, and also because their orbits bring them closer to the
Sun and to observers, where, for a given size, they are brighter
at any wavelength than their more distant relatives.  Because of their
planet-crossing orbits, the dynamical lifetimes of Centaurs are relatively
short, typically a few Myr (\eg ~{\em Horner et al.}, 2004).

The sizes of some KBOs and Centaurs have been determined by a variety
of methods. Using HST, {\em Brown and Trujillo} (2004) resolved the KBO
50000~Quaoar, placed an upper limit on the size of Sedna ({\em Brown et
al.}~2004), and resolved \Eris\ ({\em Brown et al.}, 2006). Recently
{\em Rabinowitz et al.} (2005) placed constraints on the size and albedo
of \el\ based on its short rotation period (3.9 hr) and an analysis
of the stability of a rapidly rotating ellipsoid. {\em Trilling and
Bernstein} (2006) performed a similar analysis of the lightcurves
of a number of small KBOs, obtaining constraints on their sizes and
albedos. Advances in the sensitivity of far-IR and sub-mm observatories
have recently allowed the detection of thermal emission from a sample
of outer solar system objects, providing constraints on their sizes
and albedos. {\em Jewitt et al.}~(2001), {\em Lellouch et al.}~(2002),
{\em Margot et al.}~(2002, 2004), {\em Altenhoff et al.}~(2004),
and {\em Bertoldi et al.} (2006) have reported submillimeter--millimeter
observations of thermal emission from KBOs. {\em Sykes et al.}~(1991;
1999) analyze Infrared Astronomical Satellite (IRAS) thermal detections
of 2060~Chiron and the Pluto-Charon system, determining their sizes and
albedos. Far-infrared data from the Infrared Space Observatory (ISO) were
used to determine the albedos and diameters of KBOs 15789 (1993~SC),
15874 (1996~TL$_{66}$) ({\em Thomas et al.}, 2000) and 2060~Chiron
({\em Groussin et al.}, 2004). {\em Lellouch et al.} (2000) studied the
thermal state of Pluto's surface in detail using ISO.  {\em Grundy et
al.} (2005) provide a thorough review of most of the above, and include
a sample of binary KBO systems with known masses, to constrain the
sizes and albedos of 20 KBOs.

{\em Spitzer Space Telescope} (\Spitzer\ hereafter) thermal observations
of KBOs and Centaurs have previously been reported by {\em Stansberry
et al.} (2004: 29P/Schwassmann-Wachmann 1), {\em Cruikshank et al.}
(2005: 55565 2002~AW$_{197}$), {\em Stansberry et al.} (2006: 47171
1999~TC$_{36}$), {\em Cruikshank et al.} (2006), {\em Grundy et al.}
(2007a: 65489 2003~FX$_{128}$) and {\em Grundy et al.} (in preparation: 
42355 2002~CR$_{46}$.  Here we summarize results from several \Spitzer\
programs to measure the thermal emission from 47 KBOs and Centaurs. These
observations place secure constraints on the sizes and albedos of 42 objects,
some overlapping with determinations based on other approaches mentioned
above. We present initial conclusions regarding the relationship between
albedo and orbital and physical properties of the targets, and discuss
future prospects for progress in this area.

\section{\textbf{THERMAL MODELING}}

Measurements of thermal emission can be used to constrain the sizes,
and thereby albedos, of un-resolved targets. {\em Tedesco et al.} (1992;
2002) used Infrared Astronomical Satellite (IRAS) thermal detections
of asteroids to build a catalog of albedos and diameters.  Visible
observations of the brightness of an unresolved object are inadequate to
determine its size, because that brightness is proportional to the product
of the visible geometric albedo, $p_V$, and the cross-sectional area of
the target. Similarly, the brightness in the thermal IR is proportional
to the area, and is also a function of the temperature of the surface,
which in turn depends on the albedo. Thus, measurements of both the
visible and thermal brightness can be combined to solve for both the
size of the target and its albedo. Formally the method requires the
simultaneous solution of the following two equations:
\begin{mathletters}
\begin{eqnarray}
F_{vis}&=&
                {F_{\odot, vis}\over{(r/1{\rm AU})^2}}\ R^2 p_V
                                 {{\Phi_{vis}}\over{\Delta^2}} \\
F_{ir}&=&{{R^2\Phi_{ir}}\over{\pi\Delta^2}}
    \epsilon\!\int\!B_\lambda(T(\theta,\phi))
    \sin\theta\, d\theta\, d\phi 
\end{eqnarray}
\end{mathletters}\noindent
where $F$ is the measured flux density of the object at a wavelength
in the visible (``vis'') or thermal-infrared (``ir''); $F_{\odot,
vis}$ is the visible-wavelength flux density of the Sun at 1~AU; $r$
and $\Delta$ are the object's heliocentric and geocentric distances,
respectively; $R$ is the radius of the body (assumed to be spherical);
$p_V$ is the geometric albedo in the visible; $\Phi$ is the phase
function in each regime; $B_\lambda$ is the Planck function; and
$\epsilon$ is the infrared bolometric emissivity.  $T =
T(p_V q,\eta,\epsilon,\theta,\phi)$ is the temperature, which is a
function of $p_V$; $\epsilon$; the ``beaming parameter,'' $\eta$; surface
planetographic coordinates $\theta$ and $\phi$; and the (dimensionless)
phase integral, $q$ (see below for discussions of $\eta$ and $q$).

In practice, the thermal flux depends sensitively on the temperature
distribution across the surface of the target, and uncertainties about
that temperature distribution typically dominate the uncertainties in the
derived albedos and sizes (see Fig. 1). Given knowledge of the rotation
vector, shape, and the distribution of albedo and thermal inertia,
it is in principle possible to compute the temperature
distribution. Unsurprisingly, none of these things are known for a
typical object where we seek to use the radiometric method to measure
the size and albedo.  The usual approach is to use a simplified model to
compute the temperature distribution based on little or no information
about the object's rotation axis or even rotation period.

\begin{figure*}
\epsscale{1.5}
\plotone{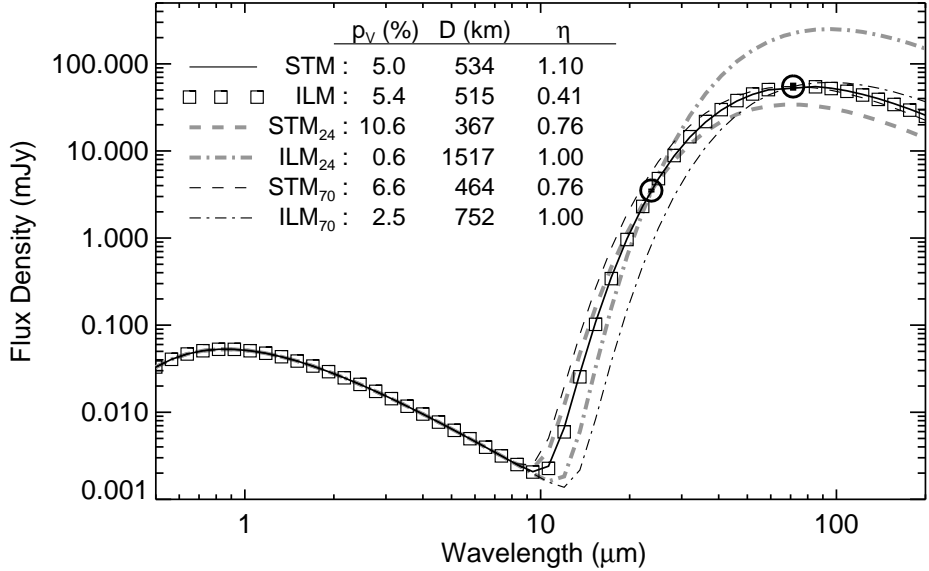}
\caption{\small Thermal models for KBO 38628 Huya (2000 EB$_{173}$).
{\em Spitzer Space Telescope} 24 and 70\mum\ data are shown as circles,
with vertical error bars within them indicating the measurement
uncertainties. Six models are fit to the data, with the resulting model
albedos, diameters, and beaming parameters summarized in the legend. From
top to bottom the models are: 1) Hybrid STM fit to 24 and 70\mum\ data,
with $\eta$ as a free parameter (the therm model used here),
2) Hybrid ILM fit to 24 and 70\mum\ data, 3) Canonical STM ($\eta=0.756$)
fit to the 24\mum\ data, 4) Canonical ILM ($\eta=1.0$) fit to the 24\mum\
data, 5) Canonical STM fit to the 70\mum\ data, 6) Canonical ILM fit to
the 70\mum\ data.  Note the close agreement of the albedos and sizes
for models 1 and 2.  Fits to data from one band, using the canonical
asteroid values for $\eta$, result in much larger uncertainties in the
derived parameters, particularly the fits to the 24\mum\ data.
}
\end{figure*}

\subsection{Standard Thermal Model}

The most commonly employed model for surface temperature on asteroidal
objects is the Standard Thermal Model (STM; cf. Lebofsky and Spencer,
1989, and references therein).  The STM assumes a non-rotating (or
equivalently, zero thermal inertia) spherical object, and represents the
``hot'' end-member to the suite of possible temperature distributions.
Under STM assumptions, the dayside temperature depends only on the
angular distance from the sub-solar point, $\theta$: $T(\theta)
= T_0 cos^{1/4}\theta$, and the temperature is zero on the night
side. The sub-solar point temperature $T_0 = [(1-A) S /
(\eta\epsilon\sigma)]^{1/4}$ . Here $A = q\, p_V$ is the bolometric
albedo, $S$ is the solar constant at the distance of the object, and
$\sigma$ is the Stefan-Boltzmann constant. Even though the STM represents
the hottest reasonable distribution of surface temperatures for an
object in radiative equilibrium with sunlight, early studies of the
emission from asteroids showed that their emission was even hotter than
predicted by the STM (Jones and Morrison, 1974; Morrison and Lebofsky,
1979). That led to the introduction of the beaming parameter, $\eta$,
which allows for localized temperature enhancements on the dayside, \eg
~in the bottoms of craters or other rough features, and the tendency
of such warm regions to radiate preferentially in the sunward (and,
for outer solar system objects, observer-ward) direction ({\it i.e.}
to beam).  (Note that while $\eta$ appears analogously to the emissivity,
$\epsilon$, in the expression for the surface temperature, $\eta$ does not
appear explicitly in the expression for the thermal emission, Eq.~1b.)
{\em Lebofsky et al.} (1986) derived a value of $\eta = 0.756$ based on
10\mum\ observations of Ceres and Pallas.  We refer to the STM with $\eta$
set to 0.756 as the {\em canonical STM}.

\subsection{Isothermal Latitude Model (ILM)}

The cold end-member of the suite of plausible temperature distributions
for an object in radiative equilibrium with sunlight is the Isothermal
Latitude Model (ILM; also known as the fast-rotator model). The ILM
assumes a spherical object illuminated at the equator and rotating
very quickly (or equivalently, a slowly rotating object with infinite
thermal inertia). The resulting temperature distribution depends only
on latitude, $\phi$: $T(\phi) = T_0 cos^{1/4}\phi$, where in this
case the sub-solar point temperature is given by $T_0 = [(1-A) S /
(\pi\eta\epsilon\sigma)]^{1/4}$.  The factor of $\pi$ in this expression
reduces the subsolar point temperature by 33\% relative to the STM.
Because the ILM is characterized by infinite thermal inertia, local
temperature variations, and therefore beaming, are precluded: thus
the {\em canonical ILM} assumes $\eta = 1$.

\subsection{A Hybrid Thermal Model}

Fig. 1 illustrates the problems inherent in using either the STM or
the ILM to measure the sizes and albedos of KBOs. In particular, none of
the 4 canonical STM or ILM models fit to either the 24 or 70\mum\ data
(4 lower elements in the figure legend) match the observed 24:70\mum\
color. As a result, the systematic uncertainties on the albedos and
diameters, depending only on whether the STM or ILM is used, are large:
$p_V$ is uncertain by a factor $>2.5$ for the fits to the 70\mum\ data, and
is uncertain by a factor of $>17$ for the fits to the 24\mum\ data.  (Note,
however, that the relative efficacy of these two wavelengths depends on
the temperature of the target: if the thermal spectrum peaks near the
24\mum\ band, observations at that wavelength will be considerably more
effective at constraining the physical properties of the target than
indicated by this particular example.)  However, if the beaming parameter,
$\eta$, is allowed to be a free parameter of the fit (top 2 elements
in the figure legend), both the color of the thermal emission and its
intensity can be matched. More importantly, both the STM and ILM give
nearly the same diameters and albedos with $\eta$ as a free parameter.
The basic reason for this is that the 24 and 70\mum\ data provide a
direct determination of the temperature of the thermal emission from
the object; equating that color temperature to the effective
temperature gives a direct estimate of the size of the target, independent
of the details of an assumed temperature distribution (and independent
of the visual brightness as well).

While the beaming parameter was introduced to model enhanced localized
dayside temperatures and infrared beaming, it can also mimic the effects
of other influences on the temperature distribution, such as pole
orientation (note that the emission from a pole-on ILM is indistinguishable
from the STM), and intermediate rotation rates and thermal inertias.
For example, a rotating body with non-zero thermal inertia will have
lower dayside temperatures than predicted by the STM, but an STM with
a value of $\eta$ larger than would be supposed based on its surface
roughness will have a similar color temperature. Likewise, 
a quickly rotating body with a low thermal inertia will have higher
dayside temperatures than predicted by the ILM, an effect that can
be mimicked by an ILM with $\eta < 1$. 

Returning to the top two models in the legend of Fig.~1, the STM fit
results in $\eta = 1.09$, suggesting that the temperature distribution
on the target (the KBO 36828~Huya) is cooler than predicted by the canonical
STM with $\eta=0.756$. Likewise, for the ILM $\eta = 0.41$, suggesting
that the surface is significantly hotter than would be predicted by the
canonical ILM with $\eta=1$.

\subsection{Thermal Model: Application}

In the following we adopt a thermal model in which the beaming
parameter, along with size and albedo, are free parameters which we use
to simultaneously fit observed flux densities at two thermal wavelengths,
and the constraint imposed by the visual brightness of the object. Because
such models have temperature distributions intermediate between the
canonical STM and ILM, they can be thought of as a hybrid between the
two. Further, because the systematic uncertainties in the model albedos
and diameters associated with the choice of hybrid STM or hybrid ILM
are fairly small relative to the uncertainties in the measured flux
densities and other model assumptions, we simply adopt the hybrid STM
as our model of choice. (The error bars on $p_V$ and $D$ stemming from
the choice of STM or ILM hybrid model in Fig.~1 are $\lesssim 4$\%
and $\lesssim 2$\%.) We note that a number of studies have employed a
similar approach with variable $\eta$ (\eg ~{\em Harris}, 1998; {\em
Delbo et al.}, 2003; {\em Fernandez et al.}, 2003).

In order to use the STM, we must make some assumptions regarding the
nature of the thermal emission and visible scattering.  We assume a gray
emissivity, $\epsilon = 0.9$. The infrared phase function, $\phi_{ir} =
0.01$\,mag/$\deg$, depends only weakly on the emission angle.  For our
observations, emission angles for all but 5 targets (29P, Asbolus, Elatus,
Thereus and Okyrhoe) were $<5\deg$. Because the effects are small relative
to other uncertainties in the models and data, we have neglected the IR
phase effect for all of the results presented here..  We assume standard
scattering behavior for the the objects in the visible, i.e. a scattering
assymetry parameter, $G = 0.15$, leading to a phase integral $q = 0.39$
({\em Bowell et al.}, 1989). This assumption also allows us to directly
relate the geometric albedo $p_V$, the diameter $D$, and the absolute
visual magnitude, $H_V$ via $D = 1346\, p_V^{1/2}\, 10^{-H_V/5}$,
where $D$ is in km ({\em Bowell et al.}, 1989; {\em Harris}, 1998).
By utilizing the absolute visual magnitude in this way, the scattering
phase function, $\Phi_{vis}$ apparently drops out; however, if the actual
scattering behavior differs from the assumption above, our albedos
and diameters will still be affected because the scattering behavior
determines the value of $q$. We note, also, the results of Romanishin
and Tegler (2005), who found that absolute magnitudes available through
the IAU Minor Planet Center and through the Horizons service at the
Jet Propulsion Laboratory have are biased downward (brighter) by 0.3
magnitudes. The $H_V$ values shown in Table~1 are culled from the
photometric literature, and should be fairly reliable.

For low albedo objects, the albedos we derive depend only weakly on the
assumed value of $q$, while for  high-albedo objects the value of $q$
exerts a strong influence (see expressions for $T_0$ in Sections~2.1
and 2.2).  For the example of 38628~Huya (Fig.~1), changing to $q =
0.8$ makes only a $\le 1$\% difference in the albedo. However, if we use
$q=0.39$ to model the data for the 4 largest objects in the sample,
\Sedna, \Eris, \el, and \fy, we obtain geometric albedos that exceed
a value of 2. While not (necessarily) unphysical, such high values for
the geometric albedo are unprecedented.  Pluto's phase integral $q= 0.8$, 
so for these 4 objects (only) we adopt that value instead.

\subsection{Thermophysical Models}

More sophisticated extensions to the STM and ILM include the effects
of surface roughness and (non-zero, non-infinite) thermal inertia
({\em Spencer}, 1990), and viewing geometries that depart significantly from
zero phase (Harris, 1998). However, for the purpose of determining KBO
albedos and diameters from their thermal emission, the hybrid STM gives
results and uncertainties that are very similar to those obtained through
application of such thermophyscal models ({\em e.g. Stansberry et al.}, 2006).
Because the hybrid STM is much simpler, and it produces results comparable
to thermophysical models, we employ only the hybrid STM.  (We note that
thermophysical models are of significant interest for objects where the
pole orientation and rotational period of the target are known, because
such models can then constrain the thermal inertia, which is of interest
in its own right).

\section{\textbf{SPITZER OBSERVATIONS}}

Roughly 310 hours of time on the \Spitzer\ have been allocated to
attempts to detect thermal emission from KBOs and Centaurs, with
the goal of measuring their albedos and diameters. \Spitzer\ has a
complement of three instruments, providing imaging capability from 3.6
-- 160\mum, and low-resolution spectroscopy from 5 -- 100\mum\ ({\em
Werner et al.}, 2004).  The long-wavelength imager, MIPS (Multiband
Imaging Photometer for \Spitzer, {\em Rieke et al.}, 2004), has 24,
70 and 160\mum\ channels. Because of the placement of these channels,
and the sensitivity of the arrays (which are at least 10 times more
sensitive than previous far-infrared satellites such as IRAS and ISO),
MIPS is well-suited to studying the thermal emission from KBOs.

\subsection{The Sample}

\Spitzer\ has targeted over 70 KBOs and Centaurs with MIPS. About 2/3 of
the observations have been succesful at detecting the thermal emission
of the target, although in some of those cases the detections have a low
signal-to-noise ratio (SNR). Here we describe observations of 47 KBOs
and Centaurs made during the first 3 years of the mission, focusing on
observations of the intrinsically brightest objects (i.e.  those with the
smallest absolute magnitudes, $H_V$), and of the Centaur objects. Table~1
summarizes the orbital and photometric properties of the sample.

The distribution of the objects in terms of dynamical class is also given,
in two forms. The second to last column, labeled ``TNO?'', indicates
whether the orbital semi-major axis is larger than Neptune's. By that
measure, 31 of the objects are trans-Neptunian Objects (TNOs), and 17 are
what might classically be called Centaur objects; that classification
is nominally in agreement with the classification scheme proposed in the
{\em Gladman et al.} chapter, although they classify Okyrhoe
and Echeclus as Jupiter family comets, rather than Centaurs. Another
classification scheme has been proposed by {\em Elliot et al.} (2005;
see also {\em Dones et al.} chapter) as a part of the Deep Ecliptic
Survey (DES) study, and the target classification thereunder appears
as the last column in the table.  According to the DES classification,
21 of the targets in the \Spitzer\ sample are Centaurs.

Thus, about 30--40\% of the sample we discuss here are Centaurs, and the
rest KBOs. Among the KBOs, only 4 objects are Classical, while 12 are in
mean-motion resonances with Neptune, 9 are in the scattered disk, and one
(\Sedna) is in the extended scattered disk: Classical objects are
under-represented. Because Classicals do not approach the Sun as closely
as the Resonant and Scattered Disk objects, and because they have somewhat
fainter absolute magnitudes, the Classicals are at the edge of \Spitzer\
capabilities.  One \Spitzer\ program has specifically targeted 15 of
the Classicals, but data analysis is ongoing.

The visible photometric properties of the sample are diverse, and
generally span the range of observed variation except in terms of the
absolute magnitudes, which for the KBOs are generally $H_V \le 7$. The
spectral properties of KBOs and Centaurs are reviewed in the chapters
by {\em Barucci et al., Tegler et al.,} and {\em Doressoundiram et al.}: here we
summarize those characteristics as regards our sample. The visible colors,
given in Table~1 as the spectral slope (measured relative to V), cover
the range from neutral to very red (Pholus).  Visible absorption features have been
reported in the 0.6--0.75\mum\ region for 47932 (2000 GN$_{171}$, 38628
Huya, and (2003 AZ$_{84}$) ({\em Lazzarin et al.}, 2003; {\em de Bergh
et al.}, 2004; {\em Fornasier et al.} 2004).  Several of the targets
exhibit near-IR spectral features, with water and methane ices being
the dominant absorbers identified.  Water ice detections have been made
for 10199 Chariklo, 83982 (2002~GO$_9$), 47171 (1999~TC$_{36}$),
47932 (2000~GN$_{171}$), 90482 Orcus, 50000 Quaoar, and \el.  55638
(2002~VE$_{95}$) exhibits methanol absorption, as does 5145~Pholus,
along with its strong water ice absorption.  Methane ice is clearly
present on \Eris\ ({\em Brown et al.} 2005), and 136472 (2005 FY$_9$)
({\em Licandro et al.}, 2006a; {\em Tegler et al.}, 2007; {\em Brown et
al.}, 2007), and \Eris\ may also have \nit\ ice ({\em Licandro et al.}
2006b). Two of the objects exhibit surface heterogeneity: 31824 Elatus
({\em Bauer et al.}, 2003) and 32532 Thereus ({\em Barucci et al.}, 2002; 
{\em Merlin et al.}, 2005).

\subsection{The Observations}

Most of the targets in the sample presented here were observed in both
the 24 and 70\mum\ channels of MIPS.  In a few cases, when the target
was predicted to be too faint to observe in the second channel, only
one channel was used. Integration times vary significantly, ranging
from 200 -- 4000~sec.  As \Spitzer\ observations of KBOs and Centaurs
proceeded, it became clear that they were significantly harder to detect
than had been predicted prior to the launch. The difficulty was due
to a combination of worse than predicted sensitivity for the 70\mum\
array (by a factor of about 2), and the fact that KBOs are colder and
smaller than assumed. As these realities made themselves evident, later
observing programs implemented more aggressive observing strategies,
and have generally been more successful than the early observations.

\begin{figure*}
\epsscale{2.0}
 \plotone{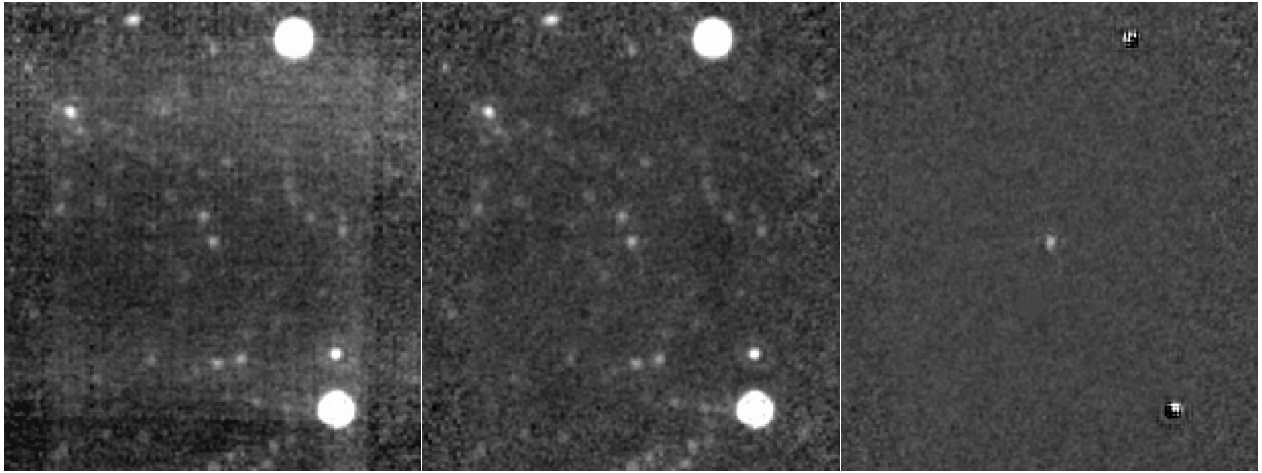}  % smaller, for astroph. not visibl in acroread, but OK on print.
%\plotone{24umProcExamp_04DW_v2.eps}  % use this one for publication.
\caption{\small Processing of the \Spitzer\ 24\mum\ data for 90482 Orcus.
The left panel shows the typical quality of image available through the
data pipeline, with scattered light and dark latent artifacts
still present. The center panel shows the improvements that can be made
by correcting the aforementioned artifacts, and reflects the quality of
the data we analyzed for targets that were imaged only once. The right
panel shows further improvement due to the subtraction of a shadow 
observation, and reflects the quality of data we analyzed for targets 
that were imaged two or more times.
}
\end{figure*}

In some cases the same target was observed more than once. These
observations fall into three categories: repeat observations seeking to
achieve higher sensitivity (\eg ~15875 (1996 TP$_{66}$) and 28978 Ixion),
multiple visits to characterize lightcurves (20000 Varuna and 47932
(2000 GN$_{171}$) are the only cases, and neither observation produced a
measurable lightcurve), and multiple visits to allow for the subtraction
of background objects (so-called ``shadow observations'').  The basic idea
of a shadow observation is to observe the target, wait for it to move out
of the way, then re-observe the field.  By subtracting the two images,
the emission from stationary sources is removed.  Fig.~2 illustrates
the shadow method, as well as some of the extra processing we apply to
the \Spitzer\ 24\mum\ data to improve its quality.

\subsection{Photometry}

Flux densities were measured using aperture photometry, as described in
{\em Cruikshank et al.} (2005) and {\em Stansberry et al.} (2006). The apertures
used encompassed the core of the PSF, out to about the first Airy
minimum (their angular radii were 10\arcsec\ and 15\arcsec\ at 24 and
70\mum). Small apertures were used to maximize the signal-to-noise
ratio (SNR) of the measurements.  Sky measurements were made in the
standard way, with an annulus surrounding the object aperture, and also
by placing multiple circular apertures in the region around the target
when the presence of background sources or cirrus structure dictated. The
photometry was aperture corrected as described in {\em Engelbracht et al.} (2007)
and {\em Gordon et al.} (2007).  Finally, we apply color corrections to our
measurements as described in {\em Stansberry et al.} (2007), resulting in
monochromatic flux densities at the effective wavelengths of the 24 and
70\mum\ filters (23.68 and 71.42\mum, respectively). The MIPS calibration
is defined in such a way that the color corrections for stellar spectra
are unity. Even though our targets are much colder (typically at or
below 80~K), the $\simeq 20$\% passbands of the MIPS filters result in
color corrections that are typically less than 10\%.

Uncertainties on the absolute calibration of MIPS are 4\% and 5\%
at 24 and 70\mum, respectively ({\em Engelbracht et al.}, 2007;
{\em Gordon et al.}, 2007).  In our photometry of KBOs and Centaurs
we adopt systematic uncertainties of 5\% and 10\%, to account for the
absolute calibration uncertainty and additional uncertainties that may
be present, \eg, in our aperture and color corrections. At 70\mum\ our
adopted systematic uncertainty includes significant margin to account for
degraded repeatability for faint sources.  Additional uncertainty comes
from the finite SNR of the detections themselves, which is estimated from
the statistics of values falling in the sky annulus and/or sky apertures.
We root-sum-square combine the systematic uncertainty with the measurement
uncertainty determined from the images to estimate the final error bars
on our measurements, and use those total uncertainties in estimating the
physical parameters we report. The SNR values we tabulate below reflect
the errors estimated from the images, and so provide an estimate of the
statistical significance of each detection.

\section{SPITZER RESULTS}

Our flux density measurements, and the albedos and diameters we derive
from them, are given in Tables~2 and 3. Table~2 gives our results for
those objects observed in both the 24 and 70\mum\ channel. When only
an upper limit on the flux density was achieved, the results in Table~2
bound the albedo and diameter of the target. Table~3 gives
the results for those objects observed at only one wavelength, and gives
a second interpretation of the data for those objects in Table~2 that
were only {\em detected} at one wavelength. 

In both Tables~2 and 3 we give the color corrected flux density of each
target, the SNR of the detections, and the temperature we used to perform
the color corrections. Where we did not detect the source, we give the
$3\sigma$ upper limit on the flux density, and the SNR column is blank.
When an object was not observed in one of the bands (Table~3 only),
the flux and SNR columns are blank. In both Table~2 and 3, albedos
($p_V$), diameters ($D$), and beaming parameters ($\eta$) follow the 
fluxes and temperatures.

\subsection{Two-Wavelength Results}

As discussed earlier and demonstrated in Fig.~1, the model-dependent
uncertainties in the albedo and diameter we derive for targets detected at
both 24 and 70\mum\ are much smaller than those uncertainties for objects
detected in only one of those bands, and in particular are usually very
much smaller than for objects detected only at 24\mum. For this reason,
we focus first on the targets we either detected at both wavelengths,
or for which we have constraints on the flux density at both. We use
these results to inform our models for targets with single-band detections
and limits.

We apply the hybrid STM to the observed flux densities as follows. For
targets {\em detected} in both bands (Table~2), we fit the observed flux densities
and the $1\sigma$ error bars, deriving albedo and diameter values
and $1\sigma$ uncertainties on them. For those objects with an {\em upper
limit} in one band and a detection in the other, we fit the detection and
the the upper limit in order to quantitatively interpret the constraints
the limit implies for the albedo and diameter.  For this second class of
observation, we also perform a single-wavelength analysis (see Table~3)
in order to derive independent constraints on these properties.  While the
results given in Table~2 include values of the beaming parameter, $\eta$,
those values only reflect the departures of the measured emission from
the assumptions of the STM; had we chosen to model the data with the ILM,
the fitted values for $\eta$ would be entirely different (even though
$p_V$ and $D$ would be very similar). Results from observations made at 
very similar epochs are averaged. An exception to that rule is the
two observations of 38628 Huya. Those data were analyzed independently to
provide a check on the repeatability of our overall data analysis and modeling 
methods for a ``bright'' KBO, and show agreement at the 4\% level for $p_V$,
and at the 2\% level for $D$.

The average behavior of the targets is of particular interest for interpreting
single-wavelelength observations, where we have no independent means for
constraining $\eta$. Restricting our attention to those targets detected
at SNR$ \ge 5$ at both 24 and 70\mum, and excluding the highest and lowest
albedo object from each class,  we find that for outer solar system
objects the average beaming parameter is $\eta = 1.2 \pm 0.35$.  We re-examine
the average properties of the sample later.

\subsection{Single-Wavelength Results}

Because we are primarily interested in the albedos and sizes of our
targets, we fit our single-wavelength observations with the STM, setting
the beaming parameter to the average value determined above: we term
this model the ``KBO-tuned'' STM.  We also apply the canonical STM and ILM
(i.e. with $\eta = 0.756$ and 1.0, respectively) to the single-wavelength
data, to interpret the data in the context of these end-member models
and assess the resulting uncertainties in model parameters.

Table~3 gives the results for the single-wavelength sample, including
those objects in Table~2 with a detection at one wavelength and an
upper-limit at the other. Where a model violates a flux limit, the
corresponding albedo and diameter entries appear as a ``?''.  The albedos
and diameters we derive using the average beaming parameter from the
two- wavelength sample are in the columns labeled ``KBO-Tuned STM'';
the range of albedos and diameters resulting from application of the
canonical STM and ILM are labeled ``STM$_0$'' and ``ILM$_0$''. Note
that the flux densities for objects in both Table~2 and 3 are sometimes
slightly different, because in Table~3 the color correction is based on
the blackbody temperature at the object's distance, rather than on the
24:70\mum\ color temperature.

\subsection{\Spitzer\ Albedos and Diameters}

The results presented above include low SNR detections, non-detections,
and multiple results for some targets. In the top portion of Table~4 we 
present results for the 39 targets that were detected at SNR $\ge 5$
at one or both wavelengths.  The results for targets that were visited
multiple times are averaged unless one observation shows some indication
of a problem. Targets with an upper limit in either band appear in both
Tables~2 and 3; in the top portion of Table~4 we give values that are
representative of all of the earlier models. The top portion of the table
contains 39 objects, 26 detected at both 24 and 70\mum, 9 at 24\mum\ only,
and 4 at 70\mum\ only. 17 of the objects have orbital semimajor axes
inside Neptune, and 21 exterior to Neptune's orbit. Where other albedo
and diameter determinations exist, the table summarizes the result,
the basis of the determination, and the publication.

\subsection{Other Constraints on $p_V$ and $D$}

The albedos and sizes of about 20 TNOs  several Centaurs have been
determined by other groups using various methods;  the lower portion of
Table~4 presents those results not given in the top portion of the table,
and the constraints that can be derived from \Spitzer\ data, when those exist
(although the SNR for all 5 cases is low, and for \Sedna\ only a 70\mum\
limit is available).

In general our results and those of other groups agree at the $\le
2\sigma$ level (\eg ~10199 Chariklo, 26308 (1998~SM$_{165}$),
47171 1999~TC$_{36}$, 55565 2002~AW$_{197}$, \Eris, \el). In a few cases
there are discrepancies. For example, our results for 20000~Varuna
are inconsistent with the millimeter results of {\em Jewitt et al.}
(2001) and {\em Lellouch et al.} (2002), which suggest a significantly larger
size and lower albedo. While our detection at 70\mum\ nominally
satisfied the $5\sigma$ threshold for Table~4, the background showed
significant structure and the SNR of the detection in the individual
visits was actually quite low. Combined with the fact that we were not
able to directly fit the beaming parameter, we are inclined to favor the
submillimeter results for this object over those from \Spitzer. While
there is some tendency for the \Spitzer\ diameters to be smaller and albedos
higher, there is generally good agreement between our \Spitzer\ results
and those from other groups and methods.

\section{ALBEDO STATISTICS and CORRELATIONS}

The Kuiper Belt is full of complexity, in terms of the dynamical history
and the spectral character of its inhabitants. It is natural to look
for relationships between the albedos of KBOs and their orbital and
other physical parameters. Fig.~3 shows the \Spitzer\ albedos for detections
with SNR$\ge 5$ (top portion of Table~4) as a function of orbital semimajor axis,
$a$, perihelion distance, $q_\sun$, object diameter, $D$, and visible
spectral slope, $S$. Because of their significant intrinsic interest,
the data for \el\ and \Sedna\ are also plotted.  Immediately apparent in
all of these plots is the marked distinction between the largest objects
(\Eris, \el, and \fy) and the rest of the objects. \Sedna\ probably
also belong to this class, although our data only place a lower bound
on its albedo.  \Eris\ and \fy\ both have abundant \met\ ice on their
surfaces, and so are expected to have very high albedos. \Sedna's near-IR
spectrum also shows evidence for \met\ and \nit\ ices ({\em Barucci et
al.}, 2005; {\em Emery et al.}, 2007), and {\em Schaller et al.} (2007)
show that those ices should not be depleted by Jean's escape: it seems
likely \Sedna's albedo is quite high. The surface of \el\ is dominated by
water ice absorptions, with no evidence for \met\ or \nit, yet also has
a very high albedo. Charon, which has a similar spectrum, has 
$p_V \simeq 37$\%, but some Saturnian satellites (notably Enceladus
and Tethys) have albedos $\ge 80$\% ({\em Morrison et al.}, 1986).

The dichotomy between \Eris, \el, \fy, Pluto (and probably \Sedna)
and the rest of the KBOs and the Centaurs, in terms of their albedos
and spectral characteristics, suggest that they are members of a unique
physical class within the Kuiper Belt population (see chapter by {\em
Brown et al.}). We will refer to these objects as ``planetoids'' in
the following, and generally exclude them from our discussion of albedo
statistics and correlations because of their obviously unique
character.

\subsection{Albedo Statistics}

Table~5 summarizes the statistics of the \Spitzer-derived albedos, and the
correlations between albedo and other parameters. Because there is no
clearly preferred way to differentiate Centaurs from KBOs, we give results
for two definitions: $a < 30.066$~AU (which we term the {\em MPC Definition},
referring to the Minor Planet Center classification (see the {\em Gladmann et al.}
chapter), and the {\em DES Definition} (referring to the Deep Ecliptic Survey 
classification ({\em Elliot et al.} 2005; {\em Dones et al.} chapter).

Typical geometric albedos for all of the KBOs and Centaurs are in the range
6.9\%--8.0\%, depending on whether the mean or median is used, with a dispersion
of about 4.1\%. Regardless of which Centaur classification one chooses, it
appears that Centaurs may have slightly lower albedos than KBOs, although
the differences are not statistically significant relative the the
dispersion of the albedos within the classes. The Kuiper variant of the
Kolmogorov-Smirnov (K-S) test gives no evidence that the albedos of the 
KBOs and Centaurs are drawn from different parent populations, regardless of 
whether the MPC or DES definition of Centaur is used. Typical values for the
beaming parameter (exluding results based on an assumed beaming parameter) 
are in the range 1.1 -- 1.20, with a dispersion of about 0.4. This is in 
good agreement with the value of $1.2 \pm 0.35$ we adopted for the ``KBO
Tuned STM'' used to construct Table~3. There does not appear to be any
significant difference in the beaming parameter between KBOs and Centaurs.

\begin{figure*}
\epsscale{1.8}
\plotone{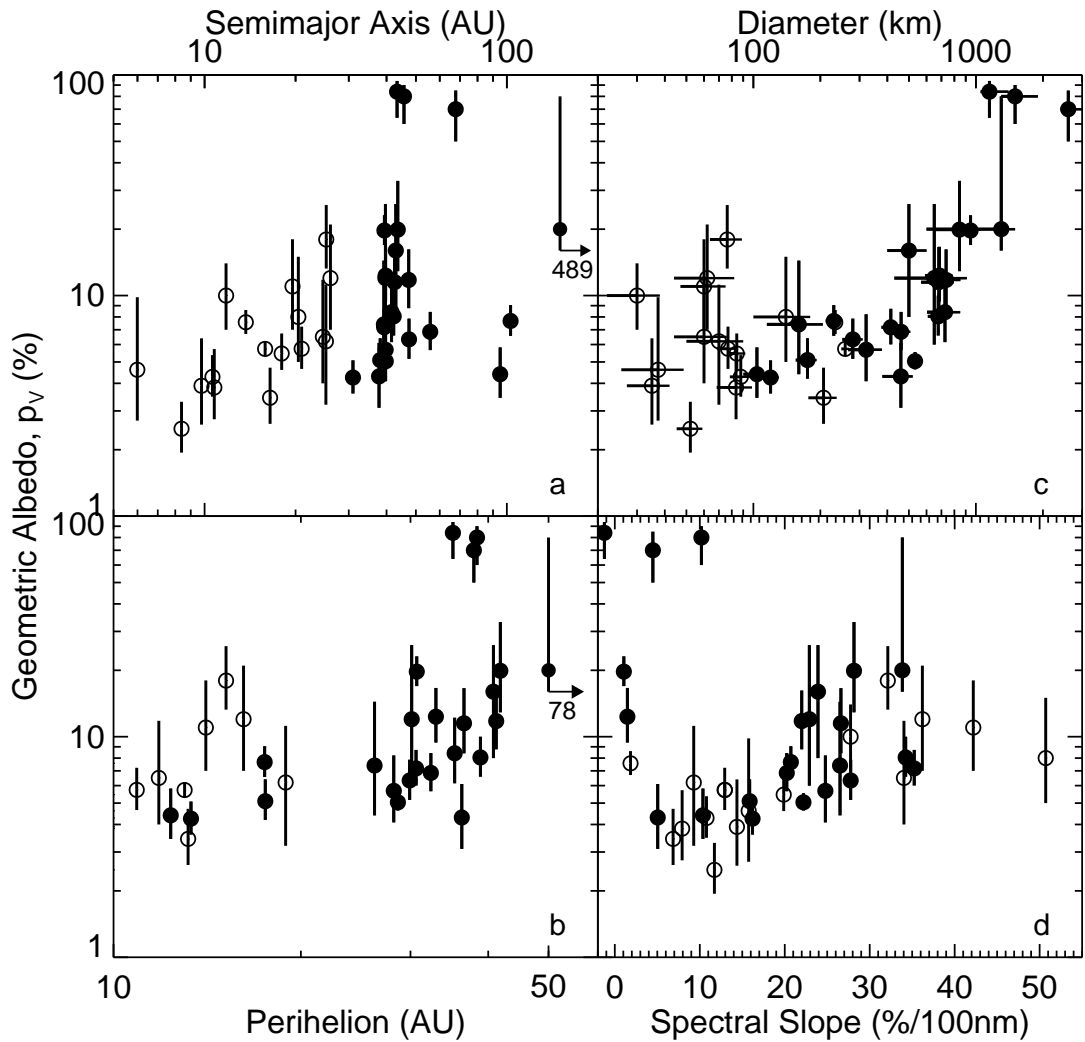}
\caption{\small Geometric albedo plotted {\em vs.} a) orbital semimajor
axis, b) orbital perihelion distance, c) object diameter, and d) the slope
of the object's visible spectrum ({\em i.e.} color). Open circles are for
Centaur objects ($a<30.066$~AU); filled circles are for TNOs. In panel a)
the point for \Sedna\ has been plotted at $a=150$~AU rather than at its
true semimajor axis of 489~AU.  In panel b) the point for \Sedna\ has
been plotted at $q=50$~AU, rather than at its true perihelion distance
of 78~AU.  
} 
\end{figure*}

\subsection{Albedo Correlations}

Because our errors are non-symmetric and probably non-gaussian, we
apply the Spearman rank-correlation test to assess the significance of
any correlations between albedo and other parameters. Table~5 gives
the correlation coefficients ($\rho$), and their significance 
($\chi$) in standard deviations from the non-correlated
case. The albedos for the 4 planetoids mentioned above are not included
in these calculations.

Fig. 3a and 3b show $p_V$ as a function of the orbital properties $a$
and $q_\sun$.  There is an upward trend of $p_V$ vs $a$, with the
objects at $a < 20$~AU clustering at $p_V \simeq 5$\%, while at larger
distances there is significantly more scatter in $p_V$. As shown in
Table~5, the correlation between $p_V$ and $a$ for the entire sample
is significant at the $\chi = 2.7\sigma$ level (99.4\% likelihood). It
appears that most of the correlation is due to the Centaurs, but the
significance of the Centaur correlation depends considerably on which
definition of Centaur is used. (Note that because the number of objects
in the KBO and Centaur subsamples is about half that of the full sample,
the significance of the correlations for the subsamples is typically
lower than that for the full sample.) Because the significance of the
$p_V$ vs $a$ correlation is below $3\sigma$, it is tentatitive. Another
reason to treat the correlation with some skepticism is that it could
reflect biases in the parameter space for KBO discoveries:
low-albedo objects will be harder to detect at visible wavelengths,
and the difficulty increases significantly with distance. Because our
sample is drawn from optically discovered objects, one might expect a
trend such as seen in Fig.~3a even if there is no real correlation
between $p_V$ and $a$.

Fig.~3b reveals a similar correlation between $p_V$ and $q_\sun$,
and Table~5 suggests that in this case the correlation is significant
at the $\chi=3.5\sigma$ level (99.95\% likelihood). This correlation
holds up fairly well for both Centaurs and KBOs, regardless of which
classification is used. It is possible that this correlation could also
be due to the discovery bias mentioned above. However, if it reflects an
actual relationship between $p_V$ and $q$, there may be a fairly simple
explanation.  Objects closer to the Sun will tend to experience higher
temperatures, depleting their surfaces of volatile molecules (which
typically have high visible reflectances).  Likewise, UV-photolosis and
Solar wind radiolysis will also proceed more quickly closer to the Sun,
and could darken those surfaces (although radiolysis by cosmic rays
probably dominates beyond about 45~AU (see {\em Cooper et al.} chapter).

Fig. 3c and 3d show $p_V$ as a function of intrinsic properties of
the objects: the diameter, $D$, and the visible spectral slope (color),
$S$. Fig.~3c shows an apparent correlation between $p_V$ and $D$, and
particularly so for the KBOs. This correlation is apparently confirmed in
Table~5, where for the MPC classification the $p_V$ vs. $D$ correlation is
significant at the $\chi = 3.4\sigma$ level (99.9\% likelihood). However,
for the DES classification the significance is only $\chi = 2.8\sigma$
(99.5\% likelihood), so the correlation is not robust against small
changes in which objects are considered as KBOs. Including the
planetoids in the correlation calculation increases the significance
of the correlation to well above $3\sigma$, but doing so results in a
(probably) false impression that the albedos of {\em all} KBOs are well
correlated with diameter. At this time it is difficult to conclude that
any such correlation exists at a statistically significant level.

Fig. 3d shows an apparent correlation between $p_V$ and $S$,
particularly for the Centaurs. Table~5 shows that this correlation is the
second most significant for a subclass, with $2.6 \le \chi \le
2.9$ (depending on the classification chosen), second only to the $p_V$
vs. $D$ correlation for KBOs. Here, the Kuiper variant K-S test does
indicate a high likelihood (99.95\%) that the albedos of red KBOs and
Centaurs (with $S>0.2$) are drawn from a different parent population than
the gray ones, a similar result to that found based on the Centaur colors
alone (see {\em Tegler et al.} chapter). A natural assumption might be
that the color diversity of KBOs and Centaurs results from mixing between
icy (bright, spectrally neutral) and organic (dark, red) components.
However, this correlation suggests that red objects systematically have
higher albedos than the gray ones.  On the basis of spectral mixing
models between spectrally neutral dark materials (such as charcoal) and
red material (represented by Titan tholin), {\em Grundy and Stansberry}
(2003) suggested that just such a correlation between red color and
higher albedo might exist. Why the Centaurs might embody this effect
more strongly than the KBOs is still a mystery. Interestingly,
the three most spectrally neutral objects defy the color--albedo trend,
having rather high albedos: there may be at least two
mechanisms underlying the observed color diversity. Those objects are 
2060~Chiron, 90482~Orcus and (2003 AZ$_{84}$), and their unique position
in the albedo-color plane may indicate that they share some unique
surface character.

\section{FUTURE PROSPECTS}

At present, \Spitzer/MIPS provides the most sensitive method available
for measuring thermal fluxes from typical KBOs, but several upcoming
observatories and instruments will provide substantially improved
sensitivity. The joint ESA/NASA Herschel mission will have at least a
factor of 2~better sensitivity at 75~microns (compared to MIPS 70~micron
sensitivity), and additionally have a number of photometry channels in the
range 70--500~microns. Since cold KBOs have their thermal emission peaks
in the range 60--100~microns, observations in the Herschel bandpasses
will map the peak of a KBO SED. Herschel is scheduled for launch in
late 2008. The Large Millimeter Telescope (LMT) in central Mexico will
have sufficient sensitivity at 1~millimeter with the SPEED instrument
to detect thermal flux from the Rayleigh-Jeans tail of cold KBOs. First
light for the LMT is expected in 2008.

Farther in the future, the American-European-Chilean Atacama Large
Millimeter Array (ALMA) will provide sufficient sensitivity from
0.35--3~millimeters to detect typical KBOs; first light for ALMA might
be as soon as 2012. The Cornell Caltech Atacama Telescope (CCAT) will operate
at 200~microns to 1~mm, and its sensitivity at 350\mum\ will
surpass that of ALMA; first light could also be in 2012. Any of Herschel,
ALMA, and CCAT (the case is less convincing for the LMT) could be used
for a large survey of many moderate-size (100~km class) KBOs. Such a
program would expand the number of KBOs with good thermal measurements
(and therefore radii and albedos) from tens to hundreds.

All of these next-generation capabilities operate at wavelengths either near
the emission peak of KBOs, or well out on the Rayleigh-Jeans part of their
spectra. While albedos and diameters derived from such observations are less
model-dependent than those based on single-wavelength observations taken
shortward of the emission peak, there are still significant uncertainties.
For example, canonical STM and ILM fits to an 850\mum\ flux density produce
albedos that differ by about 30\%; if the KBO-tuned STM is used (including
its uncertainty on $\eta$), that uncertainty is cut almost in half. If the
validity of the KBO-tuned STM is born-out by further \Spitzer\ observations
of KBOs, it can be used to significantly refine the albedos and diameters
derived from sub-millimeter KBO detections.

\section{SUMMARY}

Efforts to characterize the physical properties of KBOs and Centaurs
with \Spitzer\ are beginning to pay off. Considerable improvements have
been made in the first three years of the mission in terms of predicting
the necessary integration times, developing aggressive and successful
observing strategies, and data processing. We present our 24 and 70\mum\
observations for 47 targets (31 with orbital semimajor axes larger than
that of Neptune, 16 inside Neptune's orbit), and apply a modified version
of the Standard Thermal Model to derive albedos and diameters for them.
39 of the targets were detected at signal-to-noise ratios $\ge 5$
at one or both wavelengths. We use that sample to look for relationships
between albedo and the orbital and physical parameters of the objects. The
most marked such relationship is the distinct discontinuity in albedo
at a diameter of about 1000~km, with objects larger than that having
albedos in excess of 60\%, and those smaller than that having albedos
below about 25\%. We suggest that these large, very high albedo objects
(\Sedna, \el, \Eris and \fy) constitute a distinct class in terms of their 
physical properties.

The data suggest possible correlations of albedo with orbital distance,
and with size and color, but the statistical significance of the
correlations is marginal. Two correlations, those of albedo with
perihelion distance (for KBOs and Centaurs) and with diameter (for KBOs),
are nominally significant at more than the $3\sigma$ level.
Perhaps the most interesting trend (albeit significant at only about
the $2.8\sigma$ level) is for distinctly red Centaurs to have higher albedos 
than those that are more gray, contrary to what might intuitively be 
expected. 

Prospects for improving on and expanding these results are relatively
good.  \Spitzer\ will be operational into 2009, and more KBO observations
will probably be approved.  New ground- and space-based observatories will
also contribute significantly, and at wavelengths that are complementary
to those used here. In particular, submillimeter--millimeter studies
of KBOs should be relatively easy with facitilties such as ALMA, CCAT
and LMT.  The Herschel mission should also be very productive at far-IR
to submillimeter wavelengths.

%\textbf{ Acknowledgments.} This work was partially supported by the NASA Planetary
%Geology and Geophysics Program under grant NAG 5-10201, and
%by the National Science Foundation under grant AST-99-83530.

%\smallskip % \bigskip
%\newpage

\vskip -12pt
\centerline\textbf{ REFERENCES}
%\refs Alves, J. F., Lada, C. J., and Lada, E. A. (2001) Internal structure of a 
%cold dark molecular cloud inferred from the extinction of background 
%starlight. {\em Nature, 409}, 159-161. 

{\small
\parskip=0pt
\baselineskip=11pt

\refs Altenhoff W.J., Menten K. M. and Bertoldi F. (2002). Size dtermination of the Centaur Chariklo 
from millimeter-wavelength bolometer observations. {\em Astron. and Astrophys., 366}, L9-12.

\refs Altenhoff W. J., Bertoldi F., and Menten K. M. (2004). Size estimates of some optically
bright KBOs. {\em Astron. Astrophys., 415}, 771-775.

\refs Barucci M. A., Boehnhardt H., Dotto E., Doressoundiram A., Romon J.
{\em et al.} (2002).  Visible and near-infrared spectroscopy of the
Centaur 32532 (2001 PT$_{13}$). ESO Large Program on TNOs and Centaurs:
First spectroscopy results. {\em Astron. Astrophys., 392}, 335-339.

\refs Barucci M. A., Cruikshank D.P., Dotto E., Merlin F., Poulet F., {\em et al.} (2005).
Is Sedna another Triton? {\em Astron. Astrophys., 439}, L1-L4.  

\refs Bauer J. M., Meech K. J., Fernandez Y. R., Pittichova J., Hainaut O. R.,
Boehnhardt H. and Delsanti A. (2003). Physical survey of 24 Centaurs with visible
photometry. {\em Icarus, 166}, 195-211.

\refs Bertoldi F., Altenhoff W., Weiss A., Menten K. M.  and Thum C. (2006). The
trans-Neptunian object UB$_{313}$ is larger than Pluto. {\em Nature, 439}, 563-564.

\refs Bowell E., Hapke B., Domingue D., Lumme K., Peltoniemi J., and Harris A.
(1989).  Application of photometric models to asteroids.  In {\it Asteroids
II},  Univ.\ of Arizona Press, Tucson.

\refs Brown M. E. and Trujillo C. A. (2004).  Direct measurement of the size
of the large Kuiper belt object (50000) Quaoar. {\it Astron.\ J.,  127}, 2413-2417.

\refs Brown M. E. Trujillo C. A., Rabinowitz D., Stansberry J., Bertoldi F. and 
Koresko C. D. (2004). A Sedna update: Source, size, spectrum, surface, spin,
satellite. {\em Bull. Amer. Astron. Soc., 36}, 1068.

\refs Brown M. E., Trujillo C. A. and Rabinowitz D. L. (2005). Discovery of a
planetary-sized object in the scattered Kuiper Belt. {\em Astrophys. J., 635}, L97-L100.

\refs Brown M. E., Schaller E. L., Roe H. G., Rabinowitz D. L. and Trujillo C. A. (2006).
Direct measurement of the size of 2003 UB$_{313}$ from the Hubble Space Telescope.
{\em Astro. Phys. J., 643}, L61-63.

\refs Brown M. E., Barkume K. M., Blake G. A., Schaller E. L., Rabinowitz D. L.,
Roe H. G., Trujillo C. A. (2007). Methane and ethane on the bright Kuiper belt
object 2005 FY$_9$. {\it Astron.\ J.}, in press.

%\refs Brown R. H. Cruikshank D. P, and Pendleton Y. (1999). Water Ice on Kuiper Belt Object 1996 TO$_{66}$.
%{\em Astrophys. J., 519}, L101-L104.

\refs %Buie, M.W., D.J. Tholen, and K. Horne (1992).  Albedo maps of Pluto and
%Charon: Initial mutual event results. {\it Icarus, 97}, 211-227.

\refs Campins H., Telesco C. M., Osip D. J., Rieke G. H., Rieke M. J. and Shulz B. 
(1994). The color temperature of (2060) Chiron: a warm and small nucleus.
{\em Astron. J., 108}, 2318-2322.

\refs Cruikshank D. P. and Brown R.H. (1983). The nucleus of comet P/Schwassmann-Wachmann 1.
{\em Icarus, 56}, 377-380.

\refs Cruikshank D. P., Stansberry J. A., Emery J. P., Fernández Y. R., Werner
M. W., Trilling D. E., and Rieke G. H. (2005). The high-albedo Kuiper Belt object
(55565) 2002 AW$_{197}$. {\em Astrophys J., 624}, L53-L56.

\refs Cruikshank D. P., Barucci M. A., Emery J. P., Fernandez Y. R., Grundy W. M.,
Noll K. S. and Stansberry J. A. (2006). Physical properties of trans-Neptunian
objects. In {\em Protostars and Planets V} (Reipurth, B., Jewitt, D., Keil, K., eds).
Univ. Arizona, Tucson.

\refs Davies J., Spencer J., Sykes M., Tholen D. and Green S. (1993). (5145) Pholus.
{\em I.A.U. Circ.} 5698.

\refs de~Bergh C., Boehnhardt H., Barucci M. A., Lazzarin M., Fornasier
S., {\em et al.} (2004).  Aqueous altered silicates at the surface of
two plutinos? {\it Astron.  and Astrophys., 416}, 791-798.

\refs Delbo M., Harris A. W., Binzel R. P., Pravec P., Davies J.K. (2003). Keck
observations of near-Earth asteroids in the thermal infrared. {\em Icarus, 166}, 116-130.

%\refs Duncan M. J., and Levison H. F. (1997).  A scattered disk of icy objects and
%the origin of Jupiter-family comets. {\it Science, 276}, 1670-1672.

%\refs Elliot J. L. and Kern S. D. (2003).  Pluto's atmosphere and a
%targeted-occultation search for other bound KBO atmospheres. {\it Earth,
%Moon, and Planets, 92}, 375-393.

\refs Elliot J. L., Kern S. D., Clancy K. B., Gulbis A. A. S., Millis R. L.,
{\em et al.} (2005).  The Deep Ecliptic Survey: A search for Kuiper belt
objects and Centaurs. II. Dynamical classification, the Kuiper belt plane,
and the core population. {\it Astron.\ J., 129}, 1117-1162.

\refs Emery J. P., Dalle Ore C. M., Cruikshank D. P., Fern\'andez Y. R., Trilling D. E.
and Stansberry J. A. (2007). Ices on (90377) Sedna: Confirmation and compositional
constraints. {\em Astron. and Astrophys.}, submitted.

\refs Engelbracht C. W., Blaylock M., Su K. Y. L., Rho J., Rieke
G.H. {\em et al.} (2007).  Absolute calibration and characterization of
the Multiband Imaging Photometer for Spitzer. I. The stellar calibrator
sample and the 24 micron calibration. {\em Proc. Astron. Soc. Pacific}, submitted.

\refs Fern\'andez Y. R., Jewitt D. C., Sheppard S. S. (2002). Thermal Properties of Centaurs 
Asbolus and Chiron. {\em Astron. J., 123}, 1050-1055.

\refs Fern\'andez Y. R., Sheppard S. S., and Jewitt D. C. (2003). The Albedo Distribution of 
Jovian Trojan Asteroids. {\em Astron. J., 126}, 1563-1574.

\refs Fornasier S., Doressoundiram A., Tozzi G. P., Barucci M. A.,
Boehnhardt H.  {\em et al.} (2004).  ESO Large Program on physical
studies of trans-neptunian objects and centaurs: Final results of the
visible spectrophotometric observations. {\it Astron. and Astrophys., 421},
353-363.

\refs Gordon K. G., Engelbracht C. W., Fadda D., Stansberry J. A., Wacther S.
{\em et al.} (2007). Absolute calibration and characterization of the Multiband
Imaging Photometer for Spitzer II. 70 micron imaging. {\em Proc. Astron. Soc. Pacific}, submitted.

\refs Groussin O., Lamy P. and Jorda L. (2004). Properties of the nuclei of Centaurs
Chiron and Chariklo. {\em Astron. and Astrophys., 413}, 1163-1175.

\refs Grundy W. M. and Stansberry J. A. (2003).  Mixing models, colors, and
thermal emissions. {\it Earth, Moon, and Planets, 92}, 331-336.

\smallskip
\refs Grundy W. M., Noll K. S. and Stephens D. C. (2005).  Diverse albedos of
small trans-neptunian objects. {\it Icarus, 176}, 184-191.

\refs Grundy W. M., Stansberry J. A., Noll K. S., Stephens D. C., Trilling D. E.,
Kern S. D., Spencer J. R., Cruikshank D. P., Levison H. F. (2007). The orbit,
mass, size, albedo, and density of (65489) Ceto-Phorcys: A tidally-evolved
binary Centaur. {\em Icarus}, in press.

\refs Harris A. W. (1998). A Thermal Model for Near-Earth Asteroids.
{\em Icarus, 131}, 291-301.

\refs Horner J., Evans N. W. and Bailey M. E. (2004). Simulations of the
population of Centaurs -- I. The bulk statistics. {\em Mon. Not.
Roy. Astron. Soc., 354}, 798-810.

\refs Jewitt D. C., and Luu J. X. (1993). Discovery of the candidate Kuiper belt object 1992 QB$_{1}$.
{\em Nature, 362}, 730-732.

\refs Jewitt D. C., Aussel H. and Evans A. (2001). The size and albedo of the 
Kuiper-belt object (20000) Varuna. {\em Nature, 411}, 446-447.

\refs Jones T. J. and Morrison D. (1974).  Recalibration of the
photometric/radiometric method of determining asteroid sizes. {\it
Astron.\ J., 79}, 892-895.

\refs Lazzarin M., Barucci M. A., Boehnhardt H., Tozzi G. P., de Bergh C. and Dotto E. (2003).
ESO Large Programme on Physical Studies of Trans-Neptunian Objects and 
Centaurs: Visible Spectroscopy. {\em Astrophys. J., 125}, 1554-1558.

\refs Lebofsky L. A., Sykes M. V., Tedesco E. F., Veeder G. J., Matson D. L., {\em et al.}
(1986). A refined `standard' thermal model for asteroids based on observations 
of 1 Ceres and 2 Pallas. {\em Icarus, 68}, 239-251.

\refs Lebofsky L. A. and J. R. Spencer (1989).  Radiometry and thermal modeling of
asteroids.  In {\it Asteroids II},  Univ. of Arizona Press, Tucson.

\refs Lellouch E., Laureijs R., Schmitt B., Quirico E., de Bergh C. {\em et al.} (2000).
Pluto's non-isothermal surface. {\em Icarus, 147}, 220-250.

\refs Lellouch E., Moreno R., Ortiz J. L., Paubert G., Doressoundiram A.,
and Peixinho N. (2002). Coordinated thermal and optical observations 
of Trans-Neptunian object (20000) Varuna from Sierra Nevada. 
{\em Astron. Astrophys., 391}, 1133-1139.

\refs Levison H. F. and Duncan M. J (1997). From the Kuiper Belt to Jupiter-Family comets:
The spatial distribution of ecliptic comets. {\em Icarus, 127}, 13-32.

\refs Licandro J., Pinilla-Alonso N., Pedani M., Oliva E., Tozzi G.P. and
Grundy W. M. (2006a).  The methane ice rich surface of large TNO 2005 FY$_9$:
a Pluto-twin in the trans-neptunian belt? {\it Astron. and Astrophys., 445}, L35-L38.

\refs Licandro J., Grundy W.M., Pinilla-Alonso N. and Leysi P. (2006b).
Visible spectroscopy of TNO 2003 UB$_{313}$: Evidence for N$_2$ ice on
the surface of the largest TNO? {\it Astron. and Astrophys.} (in press).

\refs Margot J. L., Brown M. E., Trujillo C. A., and Sari R. (2002)
{\em Bull. Amer. Astron. Soc., 34}, 871 (abstract).

\refs Margot J. L., Brown M. E., Trujillo C. A., and Sari R. (2004)
{\em Bull. Amer. Astron. Soc., 36}, 1081 (abstract).

\refs Merlin F., Barucci M. A., Dotto E., de Bergh C., and Lo Curto G. (2005).
Search for surface variations on TNO 47171 and Centaur 32532.
{\em Astron. Astrophys., 444}, 977-982.

%\refs Millis R. L., Wasserman L. H., Franz O. G., Nye R. A., Elliot J. L.,
%{\em et al.} (1993).  Pluto's radius and atmosphere: Results from the entire 9 June
%1988 occultation data set. {\it Icarus, 105}, 282-297.

\refs Morrison D., Owen T. and Soderblom L. A. (1986). The satellites of Saturn,
in {\em Satellites} (J.A. Burns and M.S. Matthews, eds.), pp. 764-801.
Univ. of Arizona, Tucson.

\refs Morrison D. and Lebofsky L. A. (1979). Radiometry of asteroids, in {\em Asteroids} (T. Gehrels, ed).
Univ. Arizona, Tucson.

\refs Noll K. S., Stephens D. C., Grundy W. M., and Griffin I. (2004).
The orbit, mass, and albedo of transneptunian binary (66652) 1999 RZ$_{253}$.
{\em Icarus, 172}, 402-407.

\refs Ortiz J.L., Sota A., Moreno R., Lellouch E., Biver N. {\em et al.} (2004).
A study of Trans-Neptunian object 55636 (2002 TX$_{300}$). {\em Aston. and
Astrophys., 420}, 383-388.

\refs Osip D. J., Kern S. D., and Elliot J. L. (2003). Physical Characterization 
of the Binary Edgeworth-Kuiper Belt Object 2001 QT$_{297}$. 
{\em Earth Moon and Planets, 92}, 409-421.

\refs %Pasachoff, J.M., S.P. Souza, B.A. Babcock, D.R. Ticehurst, J.L. Elliot,
%M.J. Person, K.B. Clancy, L.C. Roberts Jr., D.T. Hall, and D.J. Tholen
%(2005).  The structure of Pluto's atmosphere from the 2002 August 21
%stellar occultation. {\it Astron.\ J., 129}, 1718-1723.

\refs Rabinowitz D. L., Barkume K., Brown M. E., Roe H., Schwartz M.,
{\em et al.} (2005).  Photometric observations constraining the size,
shape, and albedo of 2003 EL$_{61}$, a rapidly rotating, Pluto-sized object
in the Kuiper belt. {\it Astrophys.\ J., 639}, 1238-1251.

\refs Rieke G. H., Young E. T., Engelbracht C. W., Kelly D. M., Low F. J. {\em
et al.} (2004). The Multiband Imaging Photometer for Spitzer (MIPS). {\em
Astrophys. J. Suppl., 154}, 25-29.

\refs Romanishin W. and Tegler S. C. (2005). Accurate absolute magnitudes for 
Kuiper belt objects and Centaurs. {\em Icarus, 179}, 523-526.

\refs Schaller E. L., Brown M. E. (2007). Volatile loss and retention on Kuiper
Belt Objects and the depletion of nitrogen on 2005 FY$_{9}$. {\em Astrophys. J.
Lett.}, submitted.

\refs Spencer J. R. (1990).  A rough-surface thermophysical model for airless
planets. {\it Icarus, 83}, 27-38.

\refs Stansberry J. A., Van Cleve J., Reach W.T., Cruikshank D.P., Emery J.P.
{\em et al.} (2004). Spitzer Observations of the Dust Coma and Nucleus of 
29P/Schwassmann-Wachmann 1. {\em Astrophys. J. Suppl., 154}, 463-468.

\refs Stansberry J. A., Grundy W. M., Margot J. L., Cruikshank D. P., Emery J. P.
{\em et al.} (2006). The Albedo, Size, and Density of Binary Kuiper Belt 
Object (47171) 1999 TC$_{36}$. {\em Astrophys. J., 643}, 556-566.

\refs Stansberry J. A., Gordon K. D., Bhattacharya B.,  Engelbracht C. W.,
Rieke G. H., {\em et al.} (2007). Absolute calibration and characterization 
of the Multiband Imaging Photometer for Spitzer III. An asteroid-based
calibration at 160 microns. submitted to PASP.

%\refs Stern S. A., Buie M. W. and Trafton L. M. (1997).  HST high-resolution
%images and maps of Pluto. {\it Astron.\ J., 113}, 827-843.

\refs Sykes M. V. and Walker R. G. (1991). Constraints on the diameter and albedo of Chiron.
{\em Science, 251}, 777-780.

\refs Sykes M. V. (1999).  IRAS survey-mode observations of Pluto-Charon. {\it
Icarus, 142}, 155-159.

\refs Tedesco, E. F., Noah, P. V., Noah, M. and Price, S. D. (2002).  
The Supplemental IRAS Minor Planet Survey. {\em Astron. J., 123}, 1056-1085.

%\refs Tedesco E. F., Veeder G. J. Jr., Dunbar R. S. and Lebofsky L. A. (1987).
%IRAS constraints on the sizes of Pluto and Charon. {\it Nature, 327}, 127-129.

\refs Tedesco E. F., Veeder, G. J., Fowler, J. W., Chillemi, J. R. (1992).
The IRAS minor planet survey (Phillips Lab. Tech. Report PL-TR-92-2049) (Hanscom AFB, 
Massachussettes). 

\refs Tegler S. C., Grundy W. M., Romanishin W., Consolmagno G. J., Mogren K., Vilas F.
(2007).  Optical Spectroscopy of the Large Kuiper Belt Objects 136472 (2005 FY9) and 136108 
(2003 EL61). {\em Astron. J., 133}, 526-530.

\refs Thomas N., Eggers S., Ip W.-H., Lichtenberg G., Fitzsimmons A., Jorda
L., {\em et al.} (2000). Observations of the Trans-Neptunian Objects 1993 SC 
and 1996 TL$_{66}$ with the Infrared Space Observatory. {\em Astrophys. J., 534}, 446-455.

\refs Trilling D. E. and Bernstein G. M. (2006).  Light curves of 20-100 km
Kuiper belt objects using the Hubble Space Telescope. {\it Astron.\ J.,  131}, 1149-1162.

\refs Veillet C., Parker J. W., Griffin I., Marsden B., Doressoundiram A. {\em et al.}
(2002). The binary Kuiper-belt object 1998 WW31. {\em Nature, 416}, 711-713.

\refs Werner M. W., Roellig T. L., Low F. J., Rieke G. H., Rieke M. J. {\em et al.} (2004).
The Spitzer Space Telescope Mission. {\em Asrophys. J. Suppl., 154}, 1-9.

} % end \small

%} % end refernces par style

\newpage

% Table 1: Orbital and Photometric properties
% Table 1: Orbital and Photometric properties
\begin{deluxetable}{rllrrrrrrcl}
\tablecaption{Orbital and Photometric Properties\label{tbl1}}
\tablewidth{0pt}
\tabletypesize{\footnotesize}
\tablehead{
\colhead{Number\tablenotemark{a}} &
\colhead{Designation\tablenotemark{a}} &
\colhead{Name\tablenotemark{a}} &
\colhead{$a$ (AU)\tablenotemark{b}} &
\colhead{$e$\tablenotemark{b}} &
\colhead{$i$\tablenotemark{b}} &
\colhead{$H_V$\tablenotemark{c}} &
\colhead{$S$\tablenotemark{c}} &
\colhead{$\sigma_S$\tablenotemark{c}} &
\colhead{TNO?\tablenotemark{d}} &
\colhead{Class\tablenotemark{e}}
}
\startdata
    29P & \multicolumn{2}{l}{Schwassmann-Wachmann 1} 
                                     &   5.986 & 0.04 &  9.39 &  11.10 & 15.75  &  1.10  & N &  CENTR \\
   2060 & 1977 UB       & Chiron     &  13.690 & 0.38 &  6.93 &   6.58 &  1.85  &  1.18  & N &  CENTR \\
   5145 & 1992 AD       & Pholus     &  20.426 & 0.57 & 24.68 &   7.63 & 50.72  &  2.44  & N &  CENTR \\
   7066 & 1993 HA$_{2  }$ & Nessus   &  24.634 & 0.52 & 15.65 &   9.7  & 34.03  &  9.25  & N &  CENTR \\
   8405 & 1995 GO         & Asbolus  &  17.986 & 0.62 & 17.64 &   9.15 & 19.88  &  8.58  & N &  CENTR \\
  10199 & 1997 CU$_{26 }$ & Chariklo &  15.865 & 0.18 & 23.38 &   6.66 & 12.95  &  1.38  & N &  CENTR \\
  10370 & 1995 DW$_{2  }$ & Hylonome &  25.202 & 0.25 &  4.14 &   9.41 &  9.29  &  2.28  & N &  CENTR \\
  15820 & 1994 TB         &          &  39.288 & 0.31 & 12.14 &   8.00 & 40.92  &  2.87  & Y &  RESNT \\
  15874 & 1996 TL$_{66 }$ &          &  82.756 & 0.58 & 24.02 &   5.46 &  0.13  &  2.24  & Y &  SCTNR \\
  15875 & 1996 TP$_{66 }$ &          &  39.197 & 0.33 &  5.69 &   7.42 & 26.52  &  6.80  & Y &  RESNT \\
%% 19308 & 1996 TO$_{66 }$ &          &  43.177 & 0.12 & 27.48 &   4.42 &  1.13  &  1.72  & Y &  SCTNR \\
  20000 & 2000 WR$_{106}$ & Varuna   &  42.921 & 0.05 & 17.20 &   3.99 & 23.91  &  1.25  & Y &  CLSCL \\
  26308 & 1998 SM$_{165}$ &          &  47.468 & 0.37 & 13.52 &   6.38 & 27.77  &  1.91  & Y &  RESNT \\
  26375 & 1999 DE$_{9  }$ &          &  55.783 & 0.42 &  7.62 &   5.21 & 20.24  &  3.46  & Y &  RESNT \\
  28978 & 2001 KX$_{76 }$ & Ixion    &  39.648 & 0.24 & 19.59 &   3.84 & 22.90  &  1.60  & Y &  RESNT \\
  29981 & 1999 TD$_{10 }$ &          &  95.040 & 0.87 &  5.96 &   8.93 & 10.37  &  1.88  & Y &  CENTR \\
  31824 & 1999 UG$_{5  }$ & Elatus   &  11.778 & 0.38 &  5.25 &  10.52 & 27.75  &  0.97  & N &  CENTR \\
  32532 & 2001 PT$_{13 }$ & Thereus  &  10.617 & 0.20 & 20.38 &   9.32 & 10.79  &  0.96  & N &  CENTR \\
  35671 & 1998 SN$_{165}$ &          &  37.781 & 0.04 &  4.62 &   5.72 &  5.05  &  1.95  & Y &  CLSCL \\
  38628 & 2000 EB$_{173}$ & Huya     &  39.773 & 0.28 & 15.46 &   5.23 & 22.20  &  4.80  & Y &  RESNT \\
  42355 & 2002 CR$_{46 }$ & Typhon   &  38.112 & 0.54 &  2.43 &   7.65 & 15.87  &  1.93  & Y &  CENTR \\
  47171 & 1999 TC$_{36 }$ &          &  39.256 & 0.22 &  8.42 &   5.39 & 35.24  &  2.82  & Y &  RESNT \\
  47932 & 2000 GN$_{171}$ &          &  39.720 & 0.29 & 10.80 &   6.2  & 24.78  &  3.41  & Y &  RESNT \\
  50000 & 2002 LM$_{60 }$ & Quaoar   &  43.572 & 0.04 &  7.98 &   2.74 & 28.15  &  1.81  & Y &  CLSCL \\
  52872 & 1998 SG$_{35 }$ & Okyrhoe  &   8.386 & 0.31 & 15.64 &  11.04 & 11.72  &  5.08  & N &  CENTR \\
  52975 & 1998 TF$_{35 }$ & Cyllarus &  26.089 & 0.38 & 12.66 &   9.01 & 36.20  &  2.42  & N &  CENTR \\
  54598 & 2000 QC$_{243}$ & Bienor   &  16.472 & 0.20 & 20.76 &   7.70 &  6.86  &  3.17  & N &  CENTR \\
  55565 & 2002 AW$_{197}$ &          &  47.349 & 0.13 & 24.39 &   3.61 & 22.00  &  2.21  & Y &  SCTNR \\
  55576 & 2002 GB$_{10 }$ & Amycus   &  25.267 & 0.40 & 13.34 &   8.07 & 32.13  &  4.35  & N &  CENTR \\
  55636 & 2002 TX$_{300}$ &          &  43.105 & 0.12 & 25.87 &   3.49 & -0.96  &  1.20  & Y &  SCTNR \\
  55637 & 2002 UX$_{25 }$ &          &  42.524 & 0.14 & 19.48 &   3.8  & 26.61  & 10.90  & Y &  SCTNR \\
  60558 & 2000 EC$_{98 }$ & Echeclus &  10.771 & 0.46 &  4.33 &   9.55 & 10.43  &  4.83  & N &  CENTR \\
  63252 & 2001 BL$_{41 }$ &          &   9.767 & 0.29 & 12.45 &  11.47 & 14.37  &  2.75  & N &  CENTR \\
  65489 & 2003 FX$_{128}$ & Ceto     & 102.876 & 0.83 & 22.27 &   6.60 & 20.72  &  2.84  & Y &  CENTR \\
  73480 & 2002 PN$_{34 }$ &          &  30.966 & 0.57 & 16.64 &   8.66 & 16.21  &  1.90  & Y &  CENTR \\
  83982 & 2002 GO$_{9  }$ & Crantor  &  19.537 & 0.28 & 12.77 &   9.16 & 42.19  &  4.43  & N &  CENTR \\
  84522 & 2002 TC$_{302}$ &          &  55.027 & 0.29 & 35.12 &   4.1  &        &        & Y &  SCTNR \\
  84922 & 2003 VS$_{2  }$ &          &  39.273 & 0.07 & 14.79 &   4.4  &        &        & Y &  RESNT \\
  90377 & 2003 VB$_{12 }$ & Sedna    & 489.619 & 0.84 & 11.93 &   1.8  & 33.84  &  3.62  & Y &  SCEXT \\
  90482 & 2004 DW         & Orcus    &  39.363 & 0.22 & 20.59 &   2.5  &  1.06  &  1.05  & Y &  RESNT \\
  90568 & 2004 GV$_{9  }$ &          &  42.241 & 0.08 & 21.95 &   4.2  &        &        & Y &  SCTNR \\
 119951 & 2002 KX$_{14 }$ &          &  39.012 & 0.04 &  0.40 &   4.6  &        &        & Y &  CLSCL \\
 120061 & 2003 CO$_{1  }$ &          &  20.955 & 0.48 & 19.73 &   9.29 & 12.93  &  1.90  & N &  CENTR \\
 136108 & 2003 EL$_{61 }$ &          &  43.329 & 0.19 & 28.21 &   0.5  & -1.23  &  0.67  & Y &  SCTNR \\
 136199 & 2003 UB$_{313}$ & Eris     &  67.728 & 0.44 & 43.97 &  -1.1  &  4.48  &  4.63  & Y &  SCTNR \\
 136472 & 2005 FY$_{9  }$ &          &  45.678 & 0.16 & 29.00 &   0.0  & 10.19  &  2.25  & Y &  RESNT \\
        & 2002 MS$_{4  }$ &          &  41.560 & 0.15 & 17.72 &   4.0  &        &        & Y &  SCTNR \\
        & 2003 AZ$_{84 }$ &          &  39.714 & 0.17 & 13.52 &   3.71 &  1.48  &  1.01  & Y &  RESNT \\
\enddata
\tablenotetext{a}{Small body number, provisional designation, and proper name for the target sample.}
\tablenotetext{b}{Orbital semimajor axis ($a$), eccentricity ($e$) and inclination ($i$).}
\tablenotetext{c}{Absolute Visual Magnitude ($H_V$), and spectral slope and uncertainty ($S$ and $\sigma_S$,
                  in \% per 100 nm relative to V band), from the photometric literature.}
\tablenotetext{d}{Orbital semimajor axis $>$ that of Neptune (30.066~AU).}
\tablenotetext{e}{Deep Ecliptic Survey dynamical classification ({\em Elliot et al.}, 2005): CENTR = Centaur,
                  CLSCL = Classical, RESNT = Resonant, SCTNR = Scattered Near, SCEXT = Scattered Extended.}
\end{deluxetable}

%Table 2: 2-band STM Results
%Table 3: 2-band STM Results
\begin{deluxetable}{llrrrrrrrrrrrr}
\tablecaption{Two-Band Thermal Model Results\label{tbl2}} \tablewidth{0pt}
\tabletypesize{\footnotesize} 
\rotate
\tablehead{
\colhead{Number\tablenotemark{a}} &
\colhead{Name (Designation)\tablenotemark{a}} &
\colhead{AORKEY\tablenotemark{b}} &
\colhead{$R_\sun$\tablenotemark{c}} &
\colhead{$\Delta$\tablenotemark{c}} &
\colhead{$F_{24}$\tablenotemark{d}} &
\colhead{$SNR_{24}$\tablenotemark{d}} &
\colhead{$F_{70}$\tablenotemark{d}} &
\colhead{$SNR_{70}$\tablenotemark{d}} &
\colhead{$T_{24:70}$\tablenotemark{e}} &
\colhead{$p_V$\tablenotemark{f} (\%)} &
\colhead{$D$\tablenotemark{f}} &
\colhead{$\eta$\tablenotemark{f}}
}
\startdata
%                                         Color Corrected                                                                                                              Flags
%  #       Target                       AORKEY     Rsun   Delta    F24   SNR24    F70    SNR70   T_c      ---- p_V (%)----       --- Diam (km) ---        ----- eta -----     UL Fit
%============================================================================================================================================================================
   29P & Schwassmann-Wachmann      &    7864064&  5.734&  5.561& 253.783& 48.0&    96.1&  18.6&  164.7\tablenotemark{e}
                                                                                                      & $ 4.61^{+5.22}_{-1.90}$ &$  37.3^{-11.8}_{+11.3}  $&$ 0.26^{-0.18}_{+0.28}$\\[1.0pt]
  2060 & Chiron (1977 UB)          &    9033216& 13.462& 13.239&  54.410& 99.0&   145.2&  23.4&   98.1& $ 7.57^{+1.03}_{-0.87}$ &$ 233.3^{-14.4}_{+14.7}  $&$ 1.13^{-0.13}_{+0.14}$\\[1.0pt]
  5145 & Pholus (1992 AD)          &    9040896& 18.614& 18.152&   3.080& 66.0& $<$19.8&      &$>$80.2& $>6.56^{+6.38}_{-2.53}$ &$<154.5^{-44.5}_{+42.6}  $&$<1.37^{-0.48}_{+0.46}$\\[1.0pt]
  5145 & Pholus (1992 AD)          &   12661760& 19.827& 19.768&   0.962& 18.8& $<$10.1&      &$>$72.9& $>8.12^{+7.93}_{-3.17}$ &$<138.9^{-40.1}_{+38.9}  $&$<1.78^{-0.60}_{+0.57}$\\[1.0pt]
  8405 & Asbolus (1995 GO)         &    9039360&  7.743&  7.240& 202.394& 99.0&   155.7&  23.6&  141.8\tablenotemark{e}
                                                                                                      & $ 5.30^{+1.91}_{-1.25}$ &$  85.4^{-12.2}_{+12.2}  $&$ 0.66^{-0.20}_{+0.23}$\\[1.0pt]
  8405 & Asbolus (1995 GO)         &   12660480&  8.748&  8.388&  73.814& 99.0&    82.7&  11.9&  127.4& $ 5.59^{+1.69}_{-1.17}$ &$  83.2^{-10.3}_{+10.4}  $&$ 0.93^{-0.22}_{+0.25}$\\[1.0pt]
 10199 & Chariklo (1997 CU$_{26}$) &    8806144& 13.075& 12.684&  78.700& 99.0&   202.5&  24.6&   99.1& $ 5.63^{+0.76}_{-0.65}$ &$ 260.9^{-16.0}_{+16.4}  $&$ 1.17^{-0.13}_{+0.14}$\\[1.0pt]
 10199 & Chariklo (1997 CU$_{26}$) &    9038592& 13.165& 12.890&  61.509& 99.0&   177.0&  40.4&   96.3& $ 5.81^{+0.62}_{-0.55}$ &$ 256.8^{-12.8}_{+13.2}  $&$ 1.29^{-0.12}_{+0.13}$\\[1.0pt]
 10370 & Hylonome (1995 DW$_{2}$)  &    9038080& 19.963& 19.824&   0.503& 14.9& $<$10.2&      &$>$65.0& $>1.07^{+1.04}_{-0.42}$ &$<168.4^{-48.5}_{+47.3}  $&$<2.89^{-0.84}_{+0.80}$\\[1.0pt]
 15820 & (1994 TB)                 &    9042688& 28.562& 28.320&$<$0.062&     & $<$11.1&      &   48.2& $>0.55^{+0.64}_{-0.26}$ &$<451.3^{-145.4}_{+176.1}$&$ 4.87^{-1.47}_{+1.90}$\\[1.0pt]
 15874 & (1996 TL$_{66}$)          &    9035776& 35.125& 34.604&   0.380& 13.5&    22.0&   4.4&   55.6& $ 3.50^{+1.96}_{-1.07}$ &$ 575.0^{-114.6}_{+115.5}$&$ 1.76^{-0.33}_{+0.33}$\\[1.0pt]
 15875 & (1996 TP$_{66}$)          &    8805632& 26.491& 26.250&   0.689& 17.9& $<$17.6&      &$>$62.7& $>1.97^{+1.88}_{-0.76}$ &$<310.9^{-88.7}_{+86.1}  $&$<1.89^{-0.53}_{+0.50}$\\[1.0pt]
 15875 & (1996 TP$_{66}$)          &   12659456& 26.629& 26.113&   0.426& 14.6&  $<$6.9&      &$>$67.5& $>6.49^{+6.34}_{-2.54}$ &$<171.2^{-49.4}_{+48.3}  $&$<1.36^{-0.43}_{+0.41}$\\[1.0pt]
 20000 & Varuna (2000 WR$_{106}$)  &    9045760& 43.209& 42.830&$<$0.086&     &    11.0&   4.9&$<$50.1&$<11.60^{+7.66}_{-4.59}$ &$>621.2^{-139.1}_{+178.1}$&$>1.73^{-0.46}_{+0.63}$\\[1.0pt]
 26308 & (1998 SM$_{165}$)         &   14402560& 36.417& 36.087&   0.105& 15.9&     5.2&   9.4&   56.8& $ 6.33^{+1.53}_{-1.16}$ &$ 279.8^{-28.6}_{+29.7}  $&$ 1.48^{-0.17}_{+0.17}$\\[1.0pt]
 26375 & (1999 DE$_{9}$)           &    9047552& 34.980& 34.468&   0.905& 38.2&    22.6&   9.3&   62.9& $ 6.85^{+1.58}_{-1.19}$ &$ 461.0^{-45.3}_{+46.1}  $&$ 1.05^{-0.12}_{+0.12}$\\[1.0pt]
 28978 & Ixion (2001 KX$_{76}$)    &    9033472& 42.731& 42.448&   0.584& 16.6&    19.6&   3.5&   60.1& $15.65^{+12.00}_{-5.53}$&$ 573.1^{-141.9}_{+139.7}$&$ 0.82^{-0.22}_{+0.21}$\\[1.0pt]
 28978 & Ixion (2001 KX$_{76}$)    &   12659712& 42.510& 42.058&   0.290&  7.9& $<$18.4&      &$>$54.9&$>12.03^{+12.08}_{-4.89}$&$<653.6^{-191.9}_{+194.6}$&$<1.22^{-0.37}_{+0.36}$\\[1.0pt]
 29981 & (1999 TD$_{10}$)          &    8805376& 14.137& 13.945&   4.629& 31.6&    19.5&   7.2&   87.9& $ 4.40^{+1.42}_{-0.96} $&$ 103.7^{-13.5}_{+13.6}  $&$ 1.64^{-0.31}_{+0.32}$\\[1.0pt]
 31824 & Elatus (1999 UG$_{5}$)    &    9043200& 10.333&  9.998&   6.015& 69.8& $<$12.4&      &$>$105.2& $>4.86^{+5.17}_{-1.95} $&$ <47.4^{-14.4}_{+13.8}  $&$<1.46^{-0.66}_{+0.68}$\\[1.0pt]
 31824 & Elatus (1999 UG$_{5}$)    &   12661248& 11.125& 10.826&   8.596& 99.0&  $<$8.9&      &$>$118.3\tablenotemark{e}
                                                                                                      & $>9.41^{+11.57}_{-3.97}$&$ <34.1^{-11.3}_{+10.8}  $&$<0.50^{-0.29}_{+0.33}$\\[1.0pt]
 32532 & Thereus (2001 PT$_{13}$)  &    9044480&  9.813&  9.357&  25.938& 99.0&    32.7&   4.8&  122.3& $ 8.93^{+5.35}_{-2.79} $&$  60.8^{-12.7}_{+12.5}  $&$ 0.86^{-0.32}_{+0.35}$\\[1.0pt]
 32532 & Thereus (2001 PT$_{13}$)  &   12660224&  9.963&  9.685&  23.722& 99.0&    46.8&  10.3&  106.5& $ 4.28^{+1.09}_{-0.80} $&$  87.8^{-9.4}_{+9.5}    $&$ 1.50^{-0.28}_{+0.30}$\\[1.0pt]
 38628 & Huya (2000 EB$_{173}$)    &    8808192& 29.326& 29.250&   3.630& 69.4&    57.2&  10.9&   67.9& $ 4.78^{+0.94}_{-0.74} $&$ 546.5^{-47.1}_{+47.8}  $&$ 1.10^{-0.11}_{+0.12}$\\[1.0pt]
 38628 & Huya (2000 EB$_{173}$)    &    8937216& 29.325& 29.210&   3.400& 69.0&    52.9&  28.4&   68.0& $ 5.22^{+0.47}_{-0.43} $&$ 523.1^{-21.9}_{+22.7}  $&$ 1.09^{-0.07}_{+0.07}$\\[1.0pt]
 47171 & (1999 TC$_{36}$)          &    9039104& 31.098& 30.944&   1.233& 56.4&    25.3&  10.0&  64.9& $  7.18^{+1.53}_{-1.17} $&$ 414.6^{-38.2}_{+38.8}  $&$ 1.17^{-0.12}_{+0.13}$\\[1.0pt]
 47932 & (2000 GN$_{171}$)         &    9027840& 28.504& 28.009&   0.258&  8.2&    11.9&   5.6&   57.4& $ 5.68^{+2.54}_{-1.59} $&$ 321.0^{-54.2}_{+57.4}  $&$ 2.32^{-0.43}_{+0.46}$\\[1.0pt]
 50000 & Quaoar (2002 LM$_{60}$)   &   10676480& 43.345& 42.974&   0.279&  5.5&    24.6&   4.2&   52.5& $19.86^{+13.17}_{-7.04}$&$ 844.4^{-189.6}_{+206.7}$&$ 1.37^{-0.36}_{+0.39}$\\[1.0pt]
 52872 & Okyrhoe (1998 SG$_{35}$)  &    8807424&  7.793&  7.405&  28.767& 99.0&    37.4&   9.1&  121.0& $ 2.49^{+0.81}_{-0.55} $&$  52.1^{-6.9}_{+6.9}    $&$ 1.46^{-0.35}_{+0.39}$\\[1.0pt]
 54598 & Bienor (2000 QC$_{243}$)  &    9041920& 18.816& 18.350&   3.528& 78.0&    29.7&   6.1&   76.0& $ 3.44^{+1.27}_{-0.82} $&$ 206.7^{-30.1}_{+30.1}  $&$ 1.69^{-0.30}_{+0.30}$\\[1.0pt]
 55565 & (2002 AW$_{197}$)         &    9043712& 47.131& 46.701&   0.155&  7.7&    15.0&   6.7&   51.9& $11.77^{+4.42}_{-3.00} $&$ 734.6^{-108.3}_{+116.4}$&$ 1.26^{-0.20}_{+0.22}$\\[1.0pt]
 55576 & Amycus (2002 GB$_{10}$)   &   17766144& 15.589& 15.155&   6.367& 86.1&    13.6&   5.8&   99.9\tablenotemark{e}
                                                                                                      & $17.96^{+7.77}_{-4.70} $&$  76.3^{-12.5}_{+12.5}  $&$ 0.64^{-0.18}_{+0.19}$\\[1.0pt]
 55636 & (2002 TX$_{300}$)         &   10676992& 40.979& 40.729&  $<$0.065&   & $<$11.1&      &   48.4&$>17.26^{+20.33}_{-8.33}$&$<641.2^{-206.7}_{+250.3}$&$ 2.16^{-0.78}_{+0.95}$\\[1.0pt]
 55637 & (2002 UX$_{25}$)          &   10677504& 42.368& 42.413&   0.486& 15.0&    23.0&   5.3&   57.2& $11.50^{+5.09}_{-3.09} $&$ 681.2^{-114.0}_{+115.6}$&$ 1.04^{-0.18}_{+0.18}$\\[1.0pt]
 60558 & Echeclus (2000 EC$_{98}$) &    8808960& 14.141& 13.736&   4.901& 84.7&    15.5&   5.0&   94.0& $ 3.83^{+1.89}_{-1.08} $&$  83.6^{-15.2}_{+15.0}  $&$ 1.25^{-0.32}_{+0.33}$\\[1.0pt]
 65489 & Ceto (2003 FX$_{128}$)    &   17763840& 27.991& 27.674&   1.463& 71.5&    14.6&  12.2&   73.6& $ 7.67^{+1.38}_{-1.10} $&$ 229.7^{-18.2}_{+18.6}  $&$ 0.86^{-0.09}_{+0.10}$\\[1.0pt]
 73480 & (2002 PN$_{34}$)          &   17762816& 14.608& 14.153&  10.368& 99.0&    31.0&  12.6&   95.3& $ 4.25^{+0.83}_{-0.65} $&$ 119.5^{-10.2}_{+10.3}  $&$ 1.10^{-0.15}_{+0.16}$\\[1.0pt]
 83982 & Crantor (2002 GO$_{9}$)   &    9044224& 14.319& 13.824&   2.276& 58.6&  $<$8.7&      &$>$89.8& $>8.60^{+8.62}_{-3.36} $&$ <66.7^{-19.6}_{+18.7}  $&$<1.44^{-0.57}_{+0.56}$\\[1.0pt]
 84522 & (2002 TC$_{302}$)         &   13126912& 47.741& 47.654&   0.054&  6.5&    18.0&   3.1&   44.8& $ 3.08^{+2.93}_{-1.24} $&$1145.4^{-325.0}_{+337.4}$&$ 2.33^{-0.54}_{+0.53}$\\[1.0pt]
 84922 & (2003 VS$_{2}$)           &   10680064& 36.430& 36.527&   0.304&  6.0&    25.7&   3.5&   52.8& $ 5.84^{+4.78}_{-2.24} $&$ 725.2^{-187.6}_{+199.0}$&$ 2.00^{-0.51}_{+0.54}$\\[1.0pt]
 90482 & Orcus (2004 DW)           &   13000448& 47.677& 47.442&   0.329& 32.4&    26.6&  12.5&   53.1& $19.72^{+3.40}_{-2.76} $&$ 946.3^{-72.3}_{+74.1}  $&$ 1.08^{-0.09}_{+0.10}$\\[1.0pt]
 90568 & (2004 GV$_{9}$)           &   13000960& 38.992& 39.007&   0.166& 18.2&    17.5&   9.2&   51.4& $ 8.05^{+1.94}_{-1.46} $&$ 677.2^{-69.3}_{+71.3}  $&$ 1.94^{-0.20}_{+0.20}$\\[1.0pt]
119951 & (2002 KX$_{14}$)          &   10678016& 39.585& 39.197&$<$0.109&     & $<$11.7&      &   51.2& $>8.09^{+9.58}_{-3.91} $&$<561.6^{-181.5}_{+219.9}$&$ 1.91^{-0.66}_{+0.84}$\\[1.0pt]
120061 & (2003 CO$_{1}$)           &   17764864& 10.927& 10.917&  21.722& 99.0&    33.4&  11.3&  114.7& $ 5.74^{+1.49}_{-1.09} $&$  76.9^{-8.4}_{+8.5}    $&$ 0.91^{-0.18}_{+0.20}$\\[1.0pt]
136108 & (2003 EL$_{61}$)\tablenotemark{g}
                                   &   13803008& 51.244& 50.920&$<$0.022&     &     7.8&   5.3&$<$44.6&?\tablenotemark{h}~~~~   &  ?\tablenotemark{h}~~~~ & ?\tablenotemark{h}~~~~ \\[1.0pt] %!*
136199 & Eris (2003 UB$_{313}$)\tablenotemark{g}
                                   &   15909632& 96.907& 96.411&$<$0.014&     &     2.7&   4.0&$<$40.1&?\tablenotemark{h}~~~~   &  ?\tablenotemark{h}~~~~ & ?\tablenotemark{h}~~~~ \\[1.0pt] %!*
136472 & (2005 FY$_{9}$) \tablenotemark{g}
                                   &   13803776& 51.884& 51.879&   0.296& 21.1&    14.6&   9.4&  54.8&?\tablenotemark{h}~~~~   &  ?\tablenotemark{h}~~~~ & ?\tablenotemark{h}~~~~ \\[1.0pt] %!
       & (2002 MS$_{4}$)           &   10678528& 47.402& 47.488&   0.391& 20.5&    20.0&   5.1&  56.6& $ 8.41^{+3.78}_{-2.26} $&$ 726.2^{-122.9}_{+123.2}$&$ 0.88^{-0.15}_{+0.14}$\\[1.0pt]
       & (2003 AZ$_{84}$)          &   10679040& 45.669& 45.218&   0.291& 12.4&    17.8&   6.7&  55.2& $12.32^{+4.31}_{-2.91} $&$ 685.8^{-95.5}_{+98.8}  $&$ 1.04^{-0.16}_{+0.16}$\\[1.0pt]
\enddata
\tablenotetext{a}{Small body number, provisional designation, and proper name for the target sample.}
\tablenotetext{b}{Unique key identifying the data in the Spitzer data archive.}
\tablenotetext{c}{Target distance from the Sun and Spitzer, in AU.}
\tablenotetext{d}{Color-corrected flux densities (mJy) at 23.68\mum\ and 71.42\mum. Upper limits are $3\sigma$. SNR is signal to noise ratio in the images (see text).}
\tablenotetext{e}{The temperature of the blackbody spectrum used to compute the color correction. In most cases this is the 24:70 \mum\ color temperature,
                  but for the 4 denoted targets, the subsolar blackbody temperature was lower than the color temperature, and we used that instead.}
\tablenotetext{f}{The visible geometric albedo ($p_V$, percentage), diameter ($D$, km) and beaming parameter ($\eta$) from hybrid STM fits.
                  Fits to upper limits provide a quantitative interpretation of the constraints they place on $p_V$ and $D$.}
\tablenotetext{g}{Results for \Eris, \el\ and \fy\ assumed a phase integral of 0.8, typical of Pluto.}
\tablenotetext{h}{No STM with plausible albedo and beaming parameter can simultaneously fit the 24 and 70\mum\ data.
                  For 136472, models with two albedo terrains can fit the data, and give $D\simeq 1500$ km.}

\end{deluxetable}

%Table 3: 1-band STM Results
% Table 3: 1-band STM Results
%%% Was Table 4 in submitted version
\begin{deluxetable}{rlrrrrrrrrcccc}
\tablecaption{Single-Band Thermal Model Results\label{tbl3}}
\tablewidth{0pt}
% would prefer footnotesize, but the table gets run off the page if I use that
\tabletypesize{\scriptsize}  %\footnotesize} 
\rotate
\tablehead{
\colhead{} & \colhead{} &\colhead{} & \colhead{} & \colhead{} & \colhead{} & \colhead{} & \colhead{} & \colhead{} & \colhead{} &
\multicolumn{2}{c}{KBO--Tuned STM} & \colhead{$p_V$\tablenotemark{c} (\%)} & \colhead{$D$\tablenotemark{c}} \\[2pt]
\colhead{Number\tablenotemark{a}} &
\colhead{Name (Designation)\tablenotemark{a}} &
\colhead{AORKEY\tablenotemark{a}} &
\colhead{$R_\sun$\tablenotemark{a}} &
\colhead{$\Delta$\tablenotemark{a}} &
\colhead{$F_{24}$\tablenotemark{a}} &
\colhead{$SNR_{24}$\tablenotemark{a}} &
\colhead{$F_{70}$\tablenotemark{a}} &
\colhead{$SNR_{70}$\tablenotemark{a}} &
\colhead{$T_{SS}$\tablenotemark{b}} &
\colhead{$p_V$ (\%)\tablenotemark{c}} &
\colhead{$D$\tablenotemark{c}} &
\colhead{STM$_0$\ \ \ \ \ ILM$_0$} &
\colhead{STM$_0$\ \ \ \ \ ILM$_0$}
}
\startdata
%                                                              Color Corrected                               KBO-Tuned STM                    p_V (%)         Diam (km)    Flags
%  #       Target                       AORKEY    Rsun   Delta    F24  SNR24   F70    SNR70    T_0      - - - p_V (%) - - -    - - - D (km) - -       STM    ILM       STM    ILM    UL Fit
%=======================================================================================================================================================================================
   5145 & Pholus   (1992 AD)         & 9040896& 18.614& 18.152 & 3.119 &66.0 &$<$19.6&     &  91.4 &$ 8.16^{+6.16}_{- ?  } $ & $138.6^{-34.0}_{+ ?  }  $& 17.07 --  ?   &  95.8 --   ?   \\[1.0pt]% 00101
   5145 & Pholus   (1992 AD)         &12661760& 19.827& 19.768 & 0.987 &18.8 &$<$10.0&     &  88.6 &$16.18^{+11.55}_{-5.88}$ & $ 98.4^{-23.2}_{+25.0}  $& 32.74 --  ?   &  69.2 --   ?   \\[1.0pt]% 00001
   7066 & Nessus   (1993 HA$_2$)     & 9033984& 19.501& 19.219 & 0.440 &12.4 &       &     &  89.3 &$ 6.53^{+5.14}_{-2.46} $ & $ 59.7^{-15.1}_{+15.9}  $& 14.02 -- 1.44 &  40.8 -- 127.4 \\[1.0pt]% 00000
  10370 & Hylonome (1995 DW$_2$)     & 9038080& 19.963& 19.824 & 0.530 &14.9 &$<$10.0&     &  88.3 &$ 6.12^{+4.91}_{-2.33} $ & $ 70.5^{-18.0}_{+19.1}  $& 13.28 -- 1.32 &  47.9 -- 152.0 \\[1.0pt]% 00000
  10370 & Hylonome (1995 DW$_2$)     &12659968& 20.333& 20.390 & 0.451 &16.0 &       &     &  87.5 &$ 6.33^{+5.12}_{-2.42} $ & $ 69.3^{-17.8}_{+18.9}  $& 13.80 -- 1.34 &  46.9 -- 150.5 \\[1.0pt]% 00000
  15875 & (1996 TP$_{66}$)           & 8805632& 26.491& 26.250 & 0.720 &17.9 &$<$17.3&     &  76.6 &$ 5.17^{+4.98}_{-2.19} $ & $191.8^{-54.9}_{+60.9}  $& 12.54 --  ?   & 123.1 --   ?   \\[1.0pt]% 00001
  15875 & (1996 TP$_{66}$)           &12659456& 26.629& 26.113 & 0.437 &14.6 & $<$6.8&     &  76.4 &$ 8.21^{+7.61}_{- ?  } $ & $152.2^{-42.6}_{+ ?  }  $& 19.37 --  ?   &  99.1 --   ?   \\[1.0pt]% 00101
  20000 & Varuna   (2000 WR$_{106}$) & 9045760& 43.209& 42.830 &$<$0.094&    &  10.9 & 4.9 &  60.0 &$  ?                   $ & $  ?                    $&   ?   -- 8.09 &   ?   -- 744.1 \\[1.0pt]% 11110
  20000 & Varuna   (2000 WR$_{106}$) & 9031680& 43.261& 43.030 &       &     &  10.0 & 5.6 &  60.0 &$17.77^{+6.17}_{-3.79} $ & $502.0^{-69.5}_{+64.0}  $& 26.34 -- 8.68 & 412.3 -- 718.2 \\[1.0pt]% 00000
  28978 & Ixion    (2001 KX$_{76}$)  &12659712& 42.510& 42.058 & 0.303 & 7.9 &$<$18.3&     &  60.5 &$ 25.81 $\tnmd           & $ 446.3 $\tnmd           & 32.28 --  ?   & 399.1 --   ?   \\[1.0pt]% 10101
  31824 & Elatus   (1999 UG$_{5}$)   & 9043200& 10.333&  9.998 & 5.990 &69.8 &$<$12.3&     & 122.7 &$ 6.41^{+3.52}_{- ?  } $ & $ 41.3^{-8.1}_{+ ? }    $& 11.41 --  ?   &  31.0 --   ?   \\[1.0pt]% 00101
  31824 & Elatus   (1999 UG$_{5}$)   &12661248& 11.125& 10.826 & 8.596 &99.0 & $<$8.9&     & 118.3 &$  ?                   $ & $  ?                    $&   ?   --  ?   &   ?   --   ?   \\[1.0pt]% 11111
  35671 & (1998 SN$_{165}$)          & 9040384& 37.967& 37.542 &       &     &  14.7 & 6.3 &  64.0 &$ 4.33^{+1.50}_{-0.91} $ & $458.2^{-63.1}_{+57.1}  $&  6.42 -- 2.17 & 376.4 -- 648.1 \\[1.0pt]% 00000
  42355 & Typhon (2002 CR$_{46}$)    & 9029120& 17.581& 17.675 &       &     &  31.4 & 8.6 &  94.1 &$ 5.09^{+1.24}_{-0.80} $ & $173.8^{-18.0}_{+15.6}  $&  6.81 -- 3.13 & 150.3 -- 221.7 \\[1.0pt]% 00000
  52975 & Cyllarus (1998 TF$_{35}$)  & 9046528& 21.277& 21.001 & 0.274 & 8.7 &       &     &  85.5 &$11.46^{+8.96}_{-4.36} $ & $ 61.9^{-15.5}_{+16.8}  $& 24.43 -- 2.42 &  42.4 -- 134.9 \\[1.0pt]% 00000
  63252 & (2001 BL$_{41}$)           & 9032960&  9.856&  9.850 & 4.864 &95.6 &       &     & 125.7 &$ 3.90^{+2.12}_{-1.14} $ & $ 34.2^{-6.7}_{+6.5}    $&  6.93 -- 1.34 &  25.6 -- 58.3  \\[1.0pt]% 00000
  83982 & Crantor  (2002 GO$_{9}$)   & 9044224& 14.319& 13.824 & 2.310 &58.6 & $<$8.6&     & 104.2 &$11.18^{+7.09}_{- ?  } $ & $ 58.5^{-12.7}_{+ ?  }  $& 21.28 --  ?   &  42.4 --   ?   \\[1.0pt]% 00101
  90377 & Sedna    (2003 VB$_{12}$)\tablenotemark{e}\tablenotemark{f}
                                     & 8804608& 89.527& 89.291 &       &     &$<$2.4 &     &  41.7 &$>20.91^{+8.71}_{-5.29}$ &$<1268.8^{-202.7}_{+199.4}$& 32.93 -- 8.17 &1010.9 -- 2029.0\\[1.0pt]% 00000
 136108 & (2003 EL$_{61}$)\tablenotemark{e}
                                     &13803008& 51.244& 50.920 &$<$0.025&    &   7.7 & 5.3 &  55.1 &$84.11^{+9.48}_{-8.10} $ &$1151.0^{-59.9}_{+59.8}  $& 96.41 -- 59.12&1075.1 -- 1372.9\\[1.0pt]% 00000
 136199 & Eris (2003 UB$_{313}$)\tablenotemark{e}
                                     &15909632& 96.907& 96.411 &$<$0.014&    &   2.7 & 4.0 &  40.1 &$68.91^{+12.24}_{-9.98}$ &$2657.0^{-208.6}_{+216.1}$& 84.90 -- 39.17&2393.7 -- 3523.9\\[1.0pt]% 00000
 136472 &(2005 FY$_{9}$)\tablenotemark{e}
                                     &13803776& 51.884& 51.879 &\tablenotemark{g}~~
                                                                        &     &  14.6 & 9.4 &  54.8 &$78.20^{+10.30}_{-8.55}$ &$1502.9^{-90.2}_{+89.6}  $& 91.63 -- 52.55&1388.3 -- 1833.3\\[1.0pt]% 00000
 136472 &(2005 FY$_{9}$)\tablenotemark{e}
                                     &13803776& 51.884& 51.879 & 0.296 &21.1 &\tablenotemark{g}~~
                                                                                      &     &  54.8 &$35.99^{+17.56}_{-12.25}$&$2215.2^{-399.2}_{+512.4}$& 59.34 -- 6.27 &1725.3 -- 5307.0\\[1.0pt]% 00000
\enddata
\tablenotetext{a}{The first 9 columns are identical to those in Table~2. Flux densities that are blank indicate no data exist.}
\tablenotetext{b}{The subsolar temperature of a blackbody at the distance of the target.  Color corrections are made using a black 
                  body spectrum with this temperature.}
\tablenotetext{c}{The range of visible geometric albedos (given as a percentage) and diameter (in km) derived from fitting 
                  the KBO-tuned STM ({\em i.e.} $\eta=1.2\pm0.35$), and the canonical STM and ILM. ``?'' indicates the the model emission violates a flux limit.}
\tablenotetext{d}{Only the KBO-tuned STM using $\eta = 0.85$ did not violate the 70\mum\ flux limit for this observation of Ixion.}
\tablenotetext{e}{Results for \Sedna, \el, \Eris\ and \fy\ assumed a phase integral of 0.8, typical of Pluto.}
\tablenotetext{f}{Fit to the 70~\mum\ upper limit: lower bound on $p_V$, upper bound on $D$.}% Sedna note
\tablenotetext{g}{Fits to the individual bands for \fy\ are shown: it is not possible to simultaneously fit both bands with a single
                  thermal model.}
\end{deluxetable}
%
% NOTES re. models that violate the flux limits
% Added a note to indicate which models violate the 3-sigma limit for those targets with such limits.
% The numeric code at the end of each line indicates which of the models do so, with a 0 indicating
%    the model was OK relative to the limit, and 1 indicating that it violates the limit. There is
%    one digit for each model run, and the order of the models is that same as shown in the table, i.e.:
%    - first digit is for the KBO-tuned STM, nominal eta
%    - second is for KBO-tuned low-eta model (hot)
%    - third is for KBO-tuned high-eta model (cold)
%    - fourth is for the canonical STM
%    - fifth is for the canonical ILM

%Table 4: Adopted albedos and Diameters
% Table 5: Adopted albedos and diameters
\begin{deluxetable}{llccrccccc}
\tablecaption{Adopted Physical Properties\label{tbl4}}
\tablewidth{0pt}
\tabletypesize{\footnotesize}  % \footnotesize}
\rotate
\tablehead{
\colhead{} & \colhead{} & \multicolumn{4}{c}{Physical Properties from Spitzer Data} & & \multicolumn{3}{c}{Other Methods} \\
%\cline{3-7} \cline{8-9} \\
\colhead{Number\tablenotemark{a}} &
\colhead{Name (Designation)\tablenotemark{a}} &
\colhead{$p_V$\tablenotemark{a}} &
\colhead{$D$\tablenotemark{a}} &
\colhead{$\eta$\tablenotemark{a}} &
\colhead{$\lambda_{detect}$\tablenotemark{b}} &
\colhead{TNO?\tablenotemark{a}} &
\colhead{Method\tablenotemark{c}} &
\colhead{$p_V$\tablenotemark{a}} &
\colhead{$D$\tablenotemark{a}}
}
% see fit_physpars+objpars.short for input to this table
%                                                                                                                                   Other Determinations
%  #       Target                         --- p_V (%)---           -- Diam (km) --             ---- eta ----     Lambda  Class  Method   p_V (%)     D (km)
%======================================================================================================================================================
\startdata
%\\[-10pt]
%\sidehead{Objects having Spitzer detections with SNR $>$ 5}
%\multicolumn{10}{c}{Objects having Spitzer detections with SNR $>$ 5} \\[3pt]
   29P & Schwassmann-Wachmann 1    & $  4.61^{+5.22}_{-1.90}$ & $ 37.3^{ -11.8}_{+11.3}$ & $ 0.26^{-0.18}_{+0.28}$ & both & Cen & mIR  & $13\pm 4$  & $40\pm 5$\tablenotemark{Cr83} \\[0.0pt]
  2060 & Chiron (1977 UB)          & $  7.57^{+1.03}_{-0.87}$ & $233.3^{ -14.4}_{+14.7}$ & $ 1.13^{-0.13}_{+0.14}$ & both & Cen & mIR  & $17\pm 2$  &$144\pm 8$\tablenotemark{Fe02} \\[0.0pt]
       &                           &                          &                          &                         &      &     & ISO  & $11\pm 2$  &$142\pm10$\tablenotemark{Gn05} \\[0.0pt]
       &                           &                          &                          &                         &      &     & mIR  & $14\pm 5$  &    180   \tablenotemark{Ca94}  \\[0.0pt]
  5145 & Pholus (1992 AD)          & $  8.0 ^{+7   }_{-3   }$ & $140  ^{ -40  }_{+40  }$ & $ 1.3 ^{-0.4 }_{+0.4}$  &  24  & Cen & mIR  &$4.4\pm1.3$ &$189\pm26$\tablenotemark{Da93}  \\[0.0pt]
       &                           &                          &                          &                         &      &     & IRS  &$7.2\pm 2$  &$148\pm25$\tablenotemark{Cr06} \\[0.0pt]
  7066 & Nessus (1993 HA$_{2}$)    & $  6.5 ^{+5.3 }_{-2.5 }$ & $ 60  ^{ -16  }_{+16  }$ & $ 1.2 ^{-0.35}_{+0.35}$ &  24  & Cen &      &            &            \\[0.0pt]
  8405 & Asbolus (1995 GO)         & $  5.46^{+1.27}_{-0.86}$ & $ 84.2^{  -7.8}_{ +7.8}$ & $ 0.80^{-0.16}_{+0.17}$ & both & Cen & mIR  & $12\pm 3$  & $66\pm 4$\tablenotemark{Fe02} \\[0.0pt]
       &                           &                          &                          &                         &      &     & IRS  &$4.3\pm1.4$ & $95\pm 7$\tablenotemark{Cr06} \\[0.0pt]
 10199 & Chariklo (1997 CU$_{26}$) & $  5.73^{+0.49}_{-0.42}$ & $258.6^{ -10.3}_{+10.3}$ & $ 1.23^{-0.09}_{+0.10}$ & both & Cen & mm   &$5.5\pm0.5$ &    275   \tablenotemark{Al02} \\[0.0pt]
       &                           &                          &                          &                         &      &     &mIR/mm&  $7\pm 1$  &$246\pm12$\tablenotemark{Gn05} \\[0.0pt]
 10370 & Hylonome (1995 DW$_{2}$)  & $  6.2 ^{+5   }_{-3   }$ & $ 70  ^{ -20  }_{+20  }$ & $ 1.2 ^{-0.35}_{+0.35}$ &  24  & Cen &      &            &            \\[0.0pt]
 15875 & (1996 TP$_{66}$)          & $  7.4 ^{+7   }_{-3   }$ & $160  ^{ -45  }_{+45  }$ & $ 1.2 ^{-0.35}_{+0.35}$ &  24  & TNO &      &            &            \\[0.0pt]
 20000 & Varuna (2000 WR$_{106}$)  & $  16  ^{+10  }_{-8   }$ & $500  ^{ -100 }_{+100 }$ & $ 1.2 ^{-0.35}_{+0.35}$ &  70  & TNO & submm&$6\pm 2$    &$1016\pm156$\tablenotemark{Je01, Al04}\\[0.0pt]
       &                           &                          &                          &                         &      &     & mm   & $7\pm 3$   &$914\pm156$\tablenotemark{Le02, Al04} \\[0.0pt]
 26308 & (1998 SM$_{165}$)         & $  6.33^{+1.53}_{-1.16}$ & $279.8^{ -28.6}_{+29.7}$ & $ 1.48^{-0.17}_{+0.17}$ & both & TNO &mm/bin&$9.1\pm 4$  &$238\pm55$\tablenotemark{Ma04, Gy05} \\[0.0pt]
 26375 & (1999 DE$_{9}$)           & $  6.85^{+1.58}_{-1.19}$ & $461.0^{ -45.3}_{+46.1}$ & $ 1.05^{-0.12}_{+0.12}$ & both & TNO &      &            &            \\[0.0pt]
 28978 & Ixion (2001 KX$_{76}$)    & $  12 ^{+14   }_{-6   }$ &$650  ^{-220  }_{+260  }$ & $ 0.8 ^{-0.2 }_{+0.2 }$ &  24  & TNO & mm   &  $ > 15$   & $ < 804$ \tablenotemark{Al04}  \\[0.0pt]
 29981 & (1999 TD$_{10}$)          & $  4.40^{+1.42}_{-0.96}$ & $103.7^{ -13.5}_{+13.6}$ & $ 1.64^{-0.31}_{+0.32}$ & both & TNO & IRS  &  $6.5$     & $ 98 $   \tablenotemark{Cr06}  \\[0.0pt]
 31824 & Elatus (1999 UG$_{5}$)    & $  10  ^{+4   }_{-3   }$ & $ 30  ^{  -8  }_{ +8  }$ & $ 1.2 ^{-0.35}_{+0.35}$ &  24  & Cen & IRS  &$5.7\pm 2$  & $36\pm 8$\tablenotemark{Cr06}  \\[0.0pt]
 32532 & Thereus (2001 PT$_{13}$)  & $  4.28^{+1.09}_{-0.80}$ & $ 87.8^{  -9.4}_{ +9.5}$ & $ 1.50^{-0.28}_{+0.30}$ & both & Cen &      &            &            \\[0.0pt]
 35671 & (1998 SN$_{165}$)         & $  4.3 ^{+1.8 }_{-1.2 }$ & $460  ^{ -80  }_{+60  }$ & $ 1.2 ^{-0.35}_{+0.35}$ &  70  & TNO &      &            &            \\[0.0pt]
 38628 & Huya (2000 EB$_{173}$)    & $  5.04^{+0.50}_{-0.41}$ & $532.6^{ -24.4}_{+25.1}$ & $ 1.09^{-0.06}_{+0.07}$ & both & TNO & mm   &  $ > 8$    & $ < 540$ \tablenotemark{Al04}  \\[0.0pt]
 42355 & Typhon (2002 CR$_{46}$)   & $  5.1 ^{+1.3 }_{-0.9 }$ & $175  ^{ -20  }_{+17  }$ & $ 1.2 ^{-0.35}_{+0.35}$ &  70  & TNO &      &            &            \\[0.0pt]
 47171 & (1999 TC$_{36}$)          & $  7.18^{+1.53}_{-1.17}$ & $414.6^{ -38.2}_{+38.8}$ & $ 1.17^{-0.12}_{+0.13}$ & both & TNO & mm   & $5\pm 1$   &$609\pm70$\tablenotemark{Al04}  \\[0.0pt]
       &                           &                          &                          &                         &      &     &mm/bin& $14\pm 6$  &$302\pm 70$\tablenotemark{Ma04, Gy05} \\[0.0pt]
 47932 & (2000 GN$_{171}$)         & $  5.68^{+2.54}_{-1.59}$ & $321.0^{ -54.2}_{+57.4}$ & $ 2.32^{-0.43}_{+0.46}$ & both & TNO &      &            &            \\[0.0pt]
  50000 & Quaoar (2002 LM$_{60}$)  & $ 19.9^{+13.2}_{ -7.}$ & $  844 ^{  -190}_{+207}$ & $   1.4^{ -0.4}_{ +0.4}$ & both & TNO & image& $9\pm 3$   &$1260\pm190$\tablenotemark{Br04} \\[0.0pt]
 52872 & Okyrhoe (1998 SG$_{35}$)  & $  2.49^{+0.81}_{-0.55}$ & $ 52.1^{  -6.9}_{ +6.9}$ & $ 1.46^{-0.35}_{+0.39}$ & both & Cen &      &            &            \\[0.0pt]
 52975 & Cyllarus (1998 TF$_{35}$) & $ 11.5 ^{+9   }_{-5   }$ & $ 62  ^{ -18  }_{+20  }$ & $ 1.2 ^{-0.35}_{+0.35}$ &  24  & Cen &      &            &            \\[0.0pt]
 54598 & Bienor (2000 QC$_{243}$)  & $  3.44^{+1.27}_{-0.82}$ & $206.7^{ -30.1}_{+30.1}$ & $ 1.69^{-0.30}_{+0.30}$ & both & Cen &      &            &            \\[0.0pt]
 55565 & (2002 AW$_{197}$)         & $ 11.77^{+4.42}_{-3.00}$ &$734.6^{-108.3}_{+116.4}$ & $ 1.26^{-0.20}_{+0.22}$ & both & TNO & mm   & $9\pm2$    &$977\pm130$\tablenotemark{Ma02} \\[0.0pt]
 55576 & Amycus (2002 GB$_{10}$)   & $ 17.96^{+7.77}_{-4.70}$ & $ 76.3^{ -12.5}_{+12.5}$ & $ 0.64^{-0.18}_{+0.19}$ & both & Cen &      &            &            \\[0.0pt]
 55637 & (2002 UX$_{25}$)          & $ 11.50^{+5.09}_{-3.09}$ &$681.2^{-114.0}_{+115.6}$ & $ 1.04^{-0.18}_{+0.18}$ & both & TNO &      &            &            \\[0.0pt]
 60558 & Echeclus (2000 EC$_{98}$) & $  3.83^{+1.89}_{-1.08}$ & $ 83.6^{ -15.2}_{+15.0}$ & $ 1.25^{-0.32}_{+0.33}$ & both & Cen &      &            &            \\[0.0pt]
 63252 & (2001 BL$_{41}$)          & $  3.9 ^{+2.5 }_{-1.3 }$ & $ 35  ^{  -8  }_{ +7  }$ & $ 1.2 ^{-0.35}_{+0.35}$ &  24  & Cen &      &            &            \\[0.0pt]
 65489 & Ceto (2003 FX$_{128}$)    & $  7.67^{+1.38}_{-1.10}$ & $229.7^{ -18.2}_{+18.6}$ & $ 0.86^{-0.09}_{+0.10}$ & both & TNO &      &            &            \\[0.0pt]
 73480 & (2002 PN$_{34}$)          & $  4.25^{+0.83}_{-0.65}$ & $119.5^{ -10.2}_{+10.3}$ & $ 1.1 ^{-0.15}_{+0.16}$ & both & TNO &      &            &            \\[0.0pt]
 83982 & Crantor (2002 GO$_{9}$)   & $ 11   ^{+7   }_{-4   }$ & $ 60  ^{ -13  }_{+15  }$ & $ 1.20^{-0.35}_{+0.35}$ &  24  & Cen &      &            &            \\[0.0pt]
 90482 & Orcus (2004 DW)           & $ 19.72^{+3.40}_{-2.76}$ & $946.3^{ -72.3}_{+74.1}$ & $ 1.08^{-0.09}_{+0.10}$ & both & TNO &      &            &            \\[0.0pt]
 90568 & (2004 GV$_{9}$)           & $  8.05^{+1.94}_{-1.46}$ & $677.2^{ -69.3}_{+71.3}$ & $ 1.94^{-0.20}_{+0.20}$ & both & TNO &      &            &            \\[0.0pt]
120061 & (2003 CO$_{1}$)           & $  5.74^{+1.49}_{-1.09}$ & $ 76.9^{  -8.4}_{ +8.5}$ & $ 0.91^{-0.18}_{+0.20}$ & both & Cen &      &            &            \\[0.0pt]
136108 & (2003 EL$_{61}$)          & $ 84. ^{+10  }_{-20  }$ & $1150. ^{-100 }_{+250  }$ &                         &  70  & TNO &Lcurve& $65\pm 6$  &$1350\pm100$\tablenotemark{Ra05} \\[0.0pt]
136472 & (2005 FY$_{9}$)           & $ 80. ^{+10. }_{-20. }$ & $1500. ^{-200 }_{+400  }$ &                         & both & TNO &      &            &            \\[0.0pt]
       & (2002 MS$_{4}$)           & $  8.41^{+3.78}_{-2.26}$ &$726.2^{-122.9}_{+123.2}$ & $ 0.88^{-0.15}_{+0.14}$ & both & TNO &      &            &            \\[0.0pt]
       & (2003 AZ$_{84}$)          & $ 12.32^{+4.31}_{-2.91}$ & $685.8^{ -95.5}_{+98.8}$ & $ 1.04^{-0.16}_{+0.16}$ & both & TNO &      &            &            \\[3.0pt]
%%%%%%%%%%%%%%%%%%%%%%%%%%%%%%%%%%%% 
%\sidehead{Other Objects, and those having Spitzer detections with SNR $<$ 5}
%\multicolumn{10}{c}{Other Objects, and those having Spitzer detections with SNR $<$ 5} \\[0.0pt]
\tableline
  15789 & (1993 SC)                &                         &                          &                         &      & TNO & ISO  &$3.5\pm1.4$ & $298\pm140$\tablenotemark{Th00} \\[0.0pt]
  15874 & (1996 TL$_{66}$)         & $  3.5^{ +2.0}_{ -1.1}$ & $ 575 ^{  -115}_{+116 }$ & $  1.8^{ -0.3}_{ +0.3}$ & both & TNO & ISO  &  $> 1.8$   &  $< 958$   \tablenotemark{Th00} \\[0.0pt]
  19308 & (1996 TO$_{66}$)         &                         &                          &                         &      & TNO & mm   &  $> 3.3$   &  $< 902$   \tablenotemark{Al04, Gy05}\\[0.0pt]
  19521 & Chaos (1998 WH$_{24}$)   &                         &                          &                         &      & TNO & mm   &  $> 5.8$   &  $< 747$   \tablenotemark{Al04, Gy05}\\[0.0pt]
  24835 & (1995 SM$_{55}$)         &                         &                          &                         &      & TNO & mm   &  $> 6.7$   &  $< 704$   \tablenotemark{Al04, Gy05}\\[0.0pt]
  55636 & (2002 TX$_{300}$)        & $ >10                $  & $ < 800 $                &                         & limit& TNO & mm   & $ > 19$    &  $< 709$   \tablenotemark{Or04, Gy05}\\[0.0pt]
  58534 & (1997 CQ$_{29}$)         &                         &                          &                         &      & TNO & bin  & $39\pm 17$ & $77\pm 18$ \tablenotemark{Ma04, No04, Gy05}\\[0.0pt]
  66652 & (1999 RZ$_{253}$)        &                         &                          &                         &      & TNO & bin  & $29\pm 12$ & $170\pm 39$\tablenotemark{No04, Gy05}\\[0.0pt]
  84522 & (2002 TC$_{302}$)        & $  3.1^{ +2.9}_{ -1.2}$ & $1150 ^{  -325}_{+337 }$ & $  2.3^{ -0.5}_{ +0.5}$ & both & TNO & mm   &  $ >5.1$   &  $< 1211$  \tablenotemark{Al04, Gy05}\\[0.0pt]
  88611 & (2001 QT$_{297}$)        &                         &                          &                         &      & TNO & bin  & $10\pm 4$  & $168\pm 38$\tablenotemark{Os03, Gy05}\\[0.0pt]
  90377 & Sedna (2003 VB$_{12}$)   & $ >16.$                 & $<1600.$                 &                         & limit& TNO & image&  $ > 8.5$  &  $< 1800$ \tablenotemark{Br04a} \\[0.0pt]
 136199 & Eris (2003 UB$_{313}$)   & $ 70. ^{+15. }_{-20. }$ & $2600. ^{-200 }_{+400 }$ &                         &  70  & TNO & mm   & $60\pm 8$  &$3000\pm200$\tablenotemark{Be06} \\[0.0pt]
        &                          &                         &                          &                         &      &     & image& $86\pm 7$  &$2400\pm100$\tablenotemark{Br06} \\[0.0pt]
        & (1998 WW$_{31}$)         &                         &                          &                         &      & TNO & bin  & $6\pm 2.6$ & $152\pm 35$\tablenotemark{Ve02, Gy05}\\[0.0pt]
        & (2001 QC$_{298}$)        &                         &                          &                         &      & TNO & bin  &$2.5\pm 1.1$& $244\pm 55$\tablenotemark{Ma04, Gy05}\\
\enddata
\tablecomments{Results above the horizontal line have Spitzer detections with SNR $> 5$;
               those below the line have SNR $<5$, or no Spitzer data.}
\tablenotetext{a}{Columns 1--5 and 7 are as defined in Table~2.}
\tablenotetext{b}{Wavelengths where the objects were detected at SNR$>5$ (above horizontal line), or have lower quality \Spitzer\ data (below line).}
\tablenotetext{c}{''Method'' by which the diameter was measured. The meanings are:
                  ~``bin'' (binary mass plus density assumption).
                  ~``image'' (HST upper limit),
                  ~``IRS'' (Spitzer mid-IR spectra),
                  ~``ISO'' (Infrared Space Observatory),
                  ~``Lcurve'' (lightcurve + rotation dynamics),
                  ~``mIR'' (Groundbased 10--20\mum),
                  ~``mm'' (typically 1.2~mm groundbased data),
                  ~``submm'' (typically 850\mum\ groundbased data)}
\tablerefs{
(Al02)  Altenhoff et al. (2002);
(Al04)  Altenhoff et al. (2004);
(Be06)  Bertoldi et al. (2006);
(Br04)  Brown and Trujillo (2004);
(Br04a) Brown et al. (2004);
(Br06)  Brown et al. (2006);
(Ca94)  Campins et al. (1994);
(Cr83)  Cruikshank and Brown (1983);
(Cr06)  Cruikshank et al. (2006);
(Da93)  Davies et al. (1993);
(Fe02)  Fernandez et al. (2002);
(Gn05)  Groussin et al. (2004);
(Gy05)  Grundy et al. (2005);
(Je01)  Jewitt et al. (2001);
(Le01)  Lellouch et al. (2002);
(Ma02)  Margot et al. (2002);
(Ma04)  Margot et al. (2004);
(No04)  Noll et al. (2004);
(Or04)  Ortiz et al. (2004);
(Os03)  Osip et al. (2003);
(Ra05)  Rabinowitz et al. (2005);
(Th00)  Thomas et al. (2000);
(Ve02)  Veillet et al. (2002)}
\end{deluxetable}

%Table 5: Statistics and Correlations
% Table 5: Albedo Statistics and Correlations
\begin{deluxetable}{lcccccccccc}
\tablecaption{Geometric Albedo\label{tbl5}}
\tablewidth{0pt}
\tabletypesize{\small}
\tablehead{
\colhead{} & \colhead{} & \colhead{} & \multicolumn{4}{c}{MPC Classification\tablenotemark{a}} & \multicolumn{4}{c}{DES Classification\tablenotemark{b}} \\
\colhead{} & \multicolumn{2}{c}{All} & \multicolumn{2}{c}{Centaurs} & \multicolumn{2}{c}{KBOs} &\multicolumn{2}{c}{Centaurs} &\multicolumn{2}{c}{KBOs}
}
\startdata

Quantity           & \multicolumn{10}{c}{Statistics}\\
\tableline
Avgerage   & \multicolumn{2}{c}{8.01} & \multicolumn{2}{c}{6.55} & \multicolumn{2}{c}{8.87} 
                                      & \multicolumn{2}{c}{6.30} & \multicolumn{2}{c}{9.88} \\
Median     & \multicolumn{2}{c}{6.85} & \multicolumn{2}{c}{5.74} & \multicolumn{2}{c}{7.67} 
                                      & \multicolumn{2}{c}{5.73} & \multicolumn{2}{c}{8.41} \\
$\sigma$\tablenotemark{c}
           & \multicolumn{2}{c}{4.07} & \multicolumn{2}{c}{2.68} & \multicolumn{2}{c}{4.22} 
                                      & \multicolumn{2}{c}{2.50} & \multicolumn{2}{c}{4.23} \\
\# Obj.\tablenotemark{d}
           & \multicolumn{2}{c}{35}   & \multicolumn{2}{c}{15}   & \multicolumn{2}{c}{18} 
                                      & \multicolumn{2}{c}{19}   & \multicolumn{2}{c}{14} \\[6pt]
           & \multicolumn{10}{c}{Correlations}\\
Parameter       &$\rho$\tnme&$\chi$\tnmf\ \ \ \ &$\rho$\tnme&$\chi$\tnmf\ \ \ \ &$\rho$\tnme&$\chi$\tnmf\ \ \ \ &$\rho$\tnme&$\chi$\tnmf\ \ \ \ &$\rho$\tnme&$\chi$\tnmf \\
\tableline
$a$      &0.46 & 2.74\ \ \ \ & 0.70 & 2.78\ \ \ \ &0.24  &1.03\ \ \ \ &  0.41 & 1.81\ \ \ \  & 0.16 & 0.60 \\
$q_\sun$ &0.58 & 3.49\ \ \ \ & 0.58 & 2.32\ \ \ \ &0.65  &2.83\ \ \ \ &  0.43 & 1.94\ \ \ \  & 0.53 & 2.04 \\
$D$      &0.45 & 2.70\ \ \ \ &-0.08 & 0.32\ \ \ \ &0.77  &3.37\ \ \ \ & -0.07 & 0.31\ \ \ \  & 0.72 & 2.80 \\
$S$      &0.40 & 2.40\ \ \ \ & 0.64 & 2.58\ \ \ \ &0.13  &0.56\ \ \ \ &  0.66 & 2.94\ \ \ \  &-0.08 & 0.32 \\
\enddata
\tablenotetext{a}{Centaurs classified as objects having orbital semimajor axes $< 30.066$~AU.}
\tablenotetext{b}{Centaurs classified by dynamical simulations (Deep Ecliptic Survey, {\em Elliot et al.} (2005)).}
\tablenotetext{c}{Standard deviation of the albedo values.}
\tablenotetext{d}{Number of \Spitzer\ albedos used (from Table~4). The highest and lowest values were excluded.}
\tablenotetext{e}{Spearman rank correlation coefficient between albedo and the parameter in the left column.}
\tablenotetext{f}{Significance of the correlation, in standard deviations relative to the null hypothesis.}
\end{deluxetable}

\end{document}